\documentclass{aa}  
\usepackage{graphicx}
\usepackage{xcolor}
\usepackage{txfonts}
\usepackage{hyperref}
\usepackage{graphicx}
\usepackage{array}
\usepackage[export]{adjustbox} 
\newcolumntype{C}{>{\centering\arraybackslash}X}

\usepackage{siunitx}
\usepackage{gensymb}
\usepackage{placeins}
\usepackage{multirow}
\usepackage{tabularx}
\usepackage{tablefootnote}
\usepackage{array} 

\usepackage{soul}
\usepackage{color}

\hypersetup{
    colorlinks=true,
    linkcolor=blue,
    filecolor=blue,      
    urlcolor=blue,
    citecolor=blue
}
\begin{document} 
\defcitealias{hammer}{H18}
\defcitealias{Ibata2005ApJ...634..287I}{Ib05}
\defcitealias{Veljanoski2014MNRAS.442.2929V}{Vel14}
   \title{The Merger-Driven Origin of the Vast Extended Stellar Disc Around the Andromeda Galaxy}

  \titlerunning{ }
   
   \author{C. Tsakonas
          \inst{1}
          \and
          M. Arnaboldi \inst{2}
          \and 
          F. Hammer \inst{3}
          \and 
          Y. Yang \inst{3}
          \and 
          O. Gerhard \inst{4}
          \and 
          A. Burkert \inst{4,5}
          \and
          D. Hatzidimitriou \inst{1}
          }

   \institute{Department of Astrophysics, Astronomy \& Mechanics, Faculty of Physics, National and Kapodistrian University of Athens, Panepistimiopolis Zografou 15784, Greece\\
              \email{haristsak@phys.uoa.gr}
         \and
            European Southern Observatory, Karl-Schwarzschild-Str. 2, 85748 Garching, Germany
        \and
             GEPI, Observatoire de Paris, PSL Research University, CNRS, Place Jules Janssen, F-92190 Meudon, France
        \and
        Max-Planck-Institut für extraterrestrische Physik (MPE) Giessenbachstrasse 1, D-85748 Garching, Germany
        \and
         University Observatory Munich (USM) Scheinerstrasse 1, D-81679 Munich, Germany
         }

   \date{Received XX; XX}

 
  \abstract
   {The closest giant spiral, the Andromeda galaxy (M31), shows compelling evidence for a recent, gas-rich major merger event. Pronounced substructures in its inner halo and a kinematically hot stellar disc, whose star formation history shows a wide-spread star formation episode $\sim 2.5$ Gyrs ago, are telltale evidence that may be directly linked to a major (mass ratio $\sim$1:4) merger event that took place 2$-$4~Gyr ago.} 
   {Spectroscopy of resolved giant stars in the remote outskirts (R$\rm_{proj}$ = 40$-$70~kpc) of M31's disc revealed a vast extended structure that rotates with a circular velocity close to the HI gas. In addition, the spatial distribution and significant prograde rotation of two distinct, compact groups of globular clusters (GCs) in the disc outskirts (R$\rm_{proj}$~$>$~25~kpc) are unusual for typical inner halo GCs.}
   {We employ an available N-body hydrodynamical simulation of a major merger (mass ratio 1:4) that reproduces the morphology of the inner halo substructures, the age-velocity dispersion relation, and the star formation history in the disc. We compare model particles with resolved tracers (stars, GCs) in the M31 disc. To examine the evolution of the progenitor M31 disc -that appears to get stretched, distorted, and warped due to the gravitational perturbation inflicted by the major merger- we investigate the properties of the pre- versus post-merger discs of the simulated analog.}
   {The merger transforms the disc of the progenitor galaxy, which becomes kinematically hot and asymmetric. In addition, the post-merger disc gets stretched by a factor of $\sim$2 and its extent spans distances greater than 40~kpc. The stellar warp in populations older than 2~Gyr is characterized by a monotonic decrease of inclination with radius, with the outer stellar distribution appearing less edge-on at larger galactic radii.}
   {These results provide a comprehensive picture of the evolution of the giant disc of M31, the closest merger-inflicted massive galaxy. The major merger simulation successfully reproduces independent observations of resolved tracers at the outer edges of M31’s extended disc, including the presence of the warped outer disc and the misaligned angular momenta of the GCs at R$\rm_{proj}$~$>$~25~kpc. Furthermore, it provides quantitative predictions for the kinematics and surface density distributions to be tested by upcoming photometric and spectroscopic surveys.}


   \maketitle

\section{Introduction}

In the current $\Lambda$-Cold Dark Matter ($\Lambda$CDM) cosmological framework, galaxies evolve from smaller to larger entities, i.e., in a bottom-up, hierarchical fashion through smooth accretion and galaxy mergers (e.g., \citealt{WhiteRees, Bullock2005ApJ...635..931B}). Thus, mergers between dark matter halos and their baryonic content shape the observed properties of the present-day galaxies. 

Galaxy formation and evolution are systematically studied through simulations of large cosmological volumes (e.g., \citealt{Pillepich2018MNRAS.475..648P}) and their high-resolution analogs, zoom-in galactic simulations (for an extended review, see \citealt{Vogelsberger2020NatRP...2...42V} and references therein). Based on collisionless N-body models without gas, the initial consensus derived from simulated galactic encounters was that mergers between spiral galaxies typically yield an elliptical galaxy as a remnant (e.g., \citealt{Toomre1972ApJ...178..623T}). However, disc galaxies represent a major component of the morphological classification of galaxies (\citealt{Reynolds1920MNRAS..80..746R}; see also \citealt{Sandage2005ARA&A..43..581S} for the history of the morphological galaxy classification system) from high to low redshifts, while their relative fraction significantly increases toward the present epoch (e.g., \citealt{Delgado2010}). Thus, since mergers are a well-established driver of galactic evolution, it appears that disc galaxies have managed to survive their potentially destructive effects \citep{Hopkins2009ApJ...691.1168H}. 

In the last decades, the inclusion of gas in smooth particle hydrodynamical simulations (e.g., \citealt{Hernquist1989Natur.340..687H, Barnes1991ApJ...370L..65B}) and the ensuing dissipative star formation showed that gas-rich mergers may leave a disc galaxy as a merger remnant (e.g., \citealt{Barnes2002MNRAS.333..481B, Springel2005ApJ...622L...9S, Hopkins2009ApJ...691.1168H}). A young stellar disc (depending on the initial parameters) may reform from gas that retains a fraction of its angular momentum and settles in a thin rotational plane. The pre-merger, older stars will be tidally redistributed and kinematically heated. 

The Andromeda galaxy (M31) is the nearest massive spiral, sharing a comparable total mass ($\sim$~1.5~$\times$~10$^{12}$~M$_{\odot}$; see \citealt{Bhattacharya2025IAUS..379...92B} for a recent review) and a similar late-type spiral morphology with our Milky Way (MW). Its proximity ($\sim$773~kpc; \citealt{2016MNRAS.458.3282C}) allows individual stars to be resolved with 4-meter-class telescopes and followed up spectroscopically with 8-meter-class ground-based telescopes. Thus, it is the only massive spiral galaxy, apart from the MW, in which detailed investigations can be conducted on a star-by-star basis  (e.g., \citealt{Guhathakurta06, escala2020, escala22, Dey2023ApJ...944....1D}). 

In contrast to the rather quiet evolution of the MW \citep{Wyse2001ASPC..230...71W, Hammer2007, Helmi2020ARA&A..58..205H}, M31 shows compelling evidence of a more active accretion history. Conspicuous tidal debris scattered in the inner halo (e.g., \citealt{Ibata2001Natur.412...49I, Ferguson2002AJ....124.1452F, McConnachie2018ApJ...868...55M}), a notably kinematically hot thick disc \citep{2006MNRAS.369..120M, Dorman2015ApJ...803...24D, Bhattacharya2019A&A...631A..56B, Dalcanton2023AJ....166...80D}, and a widespread burst of star formation in the disc 2~$-$~3~Gyr ago \citep{Bernard2012MNRAS.420.2625B, Bernard2015MNRAS.446.2789B, Williams2017ApJ...846..145W} are clear imprints of merger/accretion events which have also impacted the chemical abundances of the younger ($< 2.5$~Gyr) and older ($4$~Gyr and older) stars \citep{Arnaboldi2022A&A...666A.109A}. Thus, M31, with an accretion history markedly different from that of the MW, serves as an archetypal merger-inflicted disc galaxy, advancing our understanding of near-field cosmology.

In addition to its halo showing several substructures, M31’s stellar and gaseous discs may provide critical insights into how mergers regulate the spatial, kinematic, and chemical evolution of the remnant galaxy. The large spectroscopic survey by \citet[][hereafter \citetalias{Ibata2005ApJ...634..287I}]{Ibata2005ApJ...634..287I} assessed the kinematic properties of the M31 disc by surveying its extreme stellar outskirts, with detections out to R$\rm_{proj}\sim$ 70~kpc. To quantify the orbital configuration of their spectroscopic sample, \citetalias{Ibata2005ApJ...634..287I} utilized a simple exponential disc model that maps the expected velocity of stars on circular orbits. Their analysis showed that stars in the remote parts of the stellar disc (R$\rm_{proj}$ up to 50$-$70~kpc) rotate with velocities close to the expected velocity of circular orbits in the plane of the M31 disc. They estimated that $\sim$10\% of the luminous mass of M31's disc resides in this extended disc component, while this region may account for $\sim$30\% of the total angular momentum of the galaxy. Thus, \citetalias{Ibata2005ApJ...634..287I} reported the presence of a vast extended disc structure around M31 and posited that its origin may be due to multiple accretion events or a single large event. 

M31 also hosts a rich system of globular clusters (GCs) with over 420 confirmed members in the Revised Bologna Catalog (RBC\footnote{https://cdsarc.cds.unistra.fr/viz-bin/cat/V/143}; \citealt{Galleti2004, Galleti2006, Galleti2007, Galleti2009a}). The vast majority of M31 GCs reside within the inner stellar disc (R$\rm_{proj}<$ 25~kpc), a region that naturally formed the focus of early M31 GC surveys (e.g., \citealt{Caldwell2011AJ....141...61C, Caldwell2016ApJ...824...42C}). The advent of deep, wide-field imaging surveys such as the Pan-Andromeda Archaeological Survey (PAndAS; \citealt{McConnachie2018ApJ...868...55M}) facilitated the comprehensive study of M31 GCs residing in the outer (R$\rm_{proj}>$ 25~kpc) parts of the galaxy. This shift marked a critical advancement in the understanding of M31’s assembly history. In contrast to the crowded inner regions, the outer halo benefits from lower stellar densities and longer dynamical timescales, increasing the likelihood that spatial and kinematic correlations between GCs and their progenitor structures are preserved. 

 \citet{Huxor2014MNRAS.442.2165H} provided the first uniform census of GCs in the M31 halo out to R$\rm_{proj}\sim$ 150~kpc. The observed spatial coincidence between many of these GCs and stellar substructures prompted suggestions of a physical connection. Based purely on their projected spatial overlap with stellar substructures, \citet{Mackey2010ApJ...717L..11M} investigated the physical connection between M31's halo GCs and underlying tidal debris. Using Bayesian statistics, they claimed that there is a very small chance that the spatial overlap of some GC groups with prominent M31 stellar substructures is due to chance alignment. This analysis was later refined by \citet{Mackey2019MNRAS.484.1756M}, who identified a dichotomy (based on the spatial overlap -or lack thereof- of GCs and stellar debris) between two distinct GC groups: GCs that are spatially associated with stellar overdensities (substructure GCs; hereafter GC-sub) and GCs with no evident overlap with inner halo substructures that are part of the smooth halo (non-substructure GCs; hereafter  GC-non). For galactic archaeology studies, GC-sub is a more relevant sample, since these GCs might have retained some memory of the original dynamical properties of their progenitor system that caused the observed spatial overlap with the underlying stars. 
 
 \citet[][hereafter \citetalias{Veljanoski2014MNRAS.442.2929V}]{Veljanoski2014MNRAS.442.2929V}  obtained velocities for 77 M31 GCs lying in R$\rm_{proj}$ = 20$-$140~kpc from the M31 centre. They reported that GCs with R$\rm_{proj}>$ 30~kpc show coherent prograde rotation with the inner stellar disc and GCs, albeit at a lower amplitude. They claim that the relatively low velocity dispersion of GC-sub suggests that this class of M31 halo GCs has been accreted along with their parent systems. \citet{Mackey2019Natur.574...69M} employed the same velocity sample and appropriate kinematic modeling to show that GC-sub and GC-non feature distinct angular momentum vectors. They posited that each of these groups may be linked to a distinct accretion event, serving as direct probes of two major accretion epochs in M31's past.

\citet[][hereafter \citetalias{hammer}]{hammer} presented a major (mass ratio $\sim$1:4) merger model between two gaseous disc galaxies, which accounts for the majority of the observational properties of the disc and inner halo of M31. The systematic analysis of the simulated M31 analogs of \citetalias{hammer} showed that a significant portion of tidally perturbed pre-merger stars from the massive progenitor fall back onto the merger remnant with an angular momentum that is misaligned relative to the inner, young disc. Meanwhile, a thin young disc emerges from cold gas deposited by the two progenitors, which retained a significant portion of its angular momentum. Shaped by collisional gas dynamics, it settles fast onto a thin orbital plane, and a subsequent burst of star formation took place. This is a well-established process in simulated wet mergers through which disc galaxies emerge after the merger of two discs (e.g., \citealt{Barnes2002MNRAS.333..481B, Springel2005ApJ...622L...9S, Hopkins2009ApJ...691.1168H}). This results in a warped and twisted stellar distribution, where the outer disc (older stars formed before the merger) exhibits a distinct orientation relative to the inner star-forming disc (younger stars, born after the merger).

 We employ the highest resolution run (model 336) of \citetalias{hammer} and ascertain that this tidal response of the pre-merger M31 galaxy to the merger-induced perturbation is a plausible physical mechanism to form the vast extended disc that encircles M31. This formation mechanism may account for the fast-rotating inner halo GCs reported in \citetalias{Veljanoski2014MNRAS.442.2929V}. We conduct direct comparisons with the kinematic, chemical, and spatial properties of the stellar populations at the edges of M31's disc to corroborate our findings. We then conclude that the available N-body model for the formation of our nearest spiral neighbor, in tandem with detailed observations, may explain the kinematic properties of our giant disc neighbor and serve as a paradigm for the formation and evolution of merger-affected disc galaxies. 

The paper is organized as follows. In Section~\ref{section: the major merger model}, we present the major merger model that underpins our study, and we quantify the geometry of the merger remnant. In Section~\ref{section:Observational properties}, we present the properties of GCs and stellar tracers at the outskirts of the M31 disc that this study aims to explain. These observed properties are compared against equivalent features of the simulated analog in Section~\ref{Section: comparison with modelled stars}. In Section~\ref{section:Structural and kinematic diagnostics}, we explore how the major merger shaped the properties of the extended disc in the model by quantifying the structural and kinematic properties of the modelled disc. Finally, in Section~\ref{section: summary and conclusions}, we summarise our findings and state the main conclusions of this work. Throughout this paper, we adopt a systemic velocity of M31 equal to V$\rm_{sys}$ = -300 km s$^ {- 1} $ \citep{Watkins2013} and an M31 distance of 773~kpc \citep{2016MNRAS.458.3282C}.

\section{The major merger model}\label{section: the major merger model}

\begin{figure*}[htp]
    \centering
    \includegraphics[width=\textwidth]{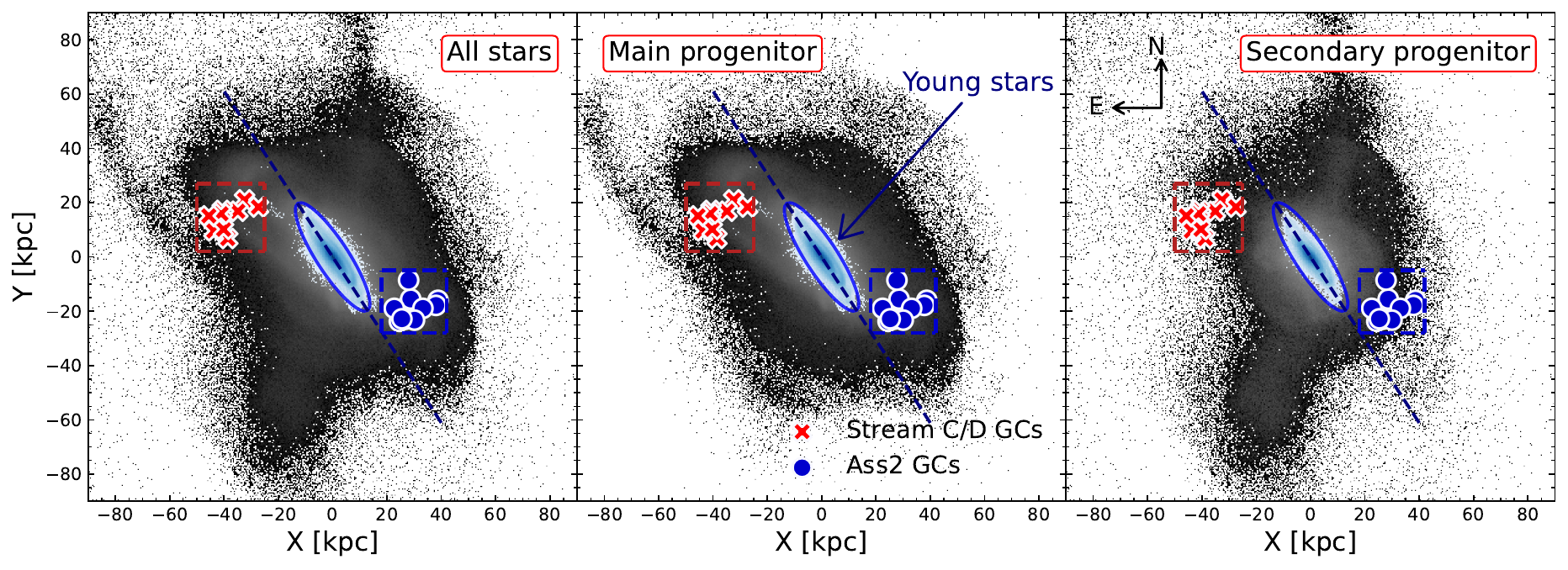}
    \caption{Spatial density map of stars in the final snapshot of the major merger model. The galaxy is rotated and projected onto the sky plane according to the inclination and PA of M31. Younger (< 2.5~Gyr) stars are plotted with a distinct color map, and their geometry is indicated by an ellipse. The panels display all particles (left), particles from the main progenitor (middle), and particles from the secondary progenitor (right). The two dominant populations of GCs in the outer M31 disc (Association 2, in the south-west, and Stream C/D in the north-east) are overlaid. Their coordinates are taken from \citetalias{Veljanoski2014MNRAS.442.2929V}. Blue open circles (red crosses) signify the blueshifted (redshifted)  kinematics in the south-west (north-east) region. Dashed boxes indicate the regions where modelled stars are selected for comparison with observations (see text).}
    \label{fig:model_gcs}
\end{figure*}

\citetalias{hammer} utilized $\sim$300 high-resolution hydrodynamical N-body simulations to reproduce M31's disc and halo properties. They ascertained a similarity with 
the observed M31's inner structures 
and the substructures in its inner halo (Giant Stellar Stream, the North-East and Western shelves) which can be associated with a single accretion event. A major merger between two gaseous disc galaxies simultaneously reproduces the kinematically hot thick disc of M31 observed at the present time (\citealt{2006MNRAS.369..120M, Dorman2015ApJ...803...24D, paperII}), the widespread episode of star formation in the disc 2$-$3~Gyr ago (\citealt{Bernard2012MNRAS.420.2625B, Bernard2015MNRAS.446.2789B, Dalcanton2012, Williams2017ApJ...846..145W}), and the intricate morphological and kinematical characteristics of M31's inner halo \citep{paperVI, Tsakonas2025A&A...699A..56T} including the Giant Stellar Stream. 

We employ model~336 from \citetalias{hammer}, as it is the one that most accurately reproduced the observations of M31 (e.g., the radial profile of the stellar surface density of the inner halo, the morphology of the substructures, etc.) at the time it was published. This model was selected by \citetalias{hammer} for their highest-resolution run with 20 million particles. It comprises two progenitor galaxies, each containing co-planar stellar and gaseous discs, both assumed to be thin (i.e., with a scale height of 1/10 of their scale length). Each gas disc is three times more extended than its respective stellar disc. The gas fractions are 50\% and 80\% of their baryonic content, for the main and the secondary progenitor, respectively. The dark matter content of each galaxy is 80$\%$ of its total mass.

The simulation begins at a lookback time of 8~Gyr, with the two discs initially separated by $\sim$500~kpc. The first pericentre passage of the secondary galaxy occurs 1$-$2 Gyr afterwards. It has a moderate impact on the massive M31 progenitor. After the second pericentre passage, at a lookback time of 2$-$4 Gyr, the two progenitors begin to merge into a single remnant.  In the final snapshot, the central parts of the secondary are completely dissolved; thus, a residual core is absent from the remnant galaxy.

\subsection{The spatial configuration of the merger remnant}\label{section: Spatial configuration of the remnant}

\begin{figure*}[htp]
    \centering
    \includegraphics[width=\textwidth]{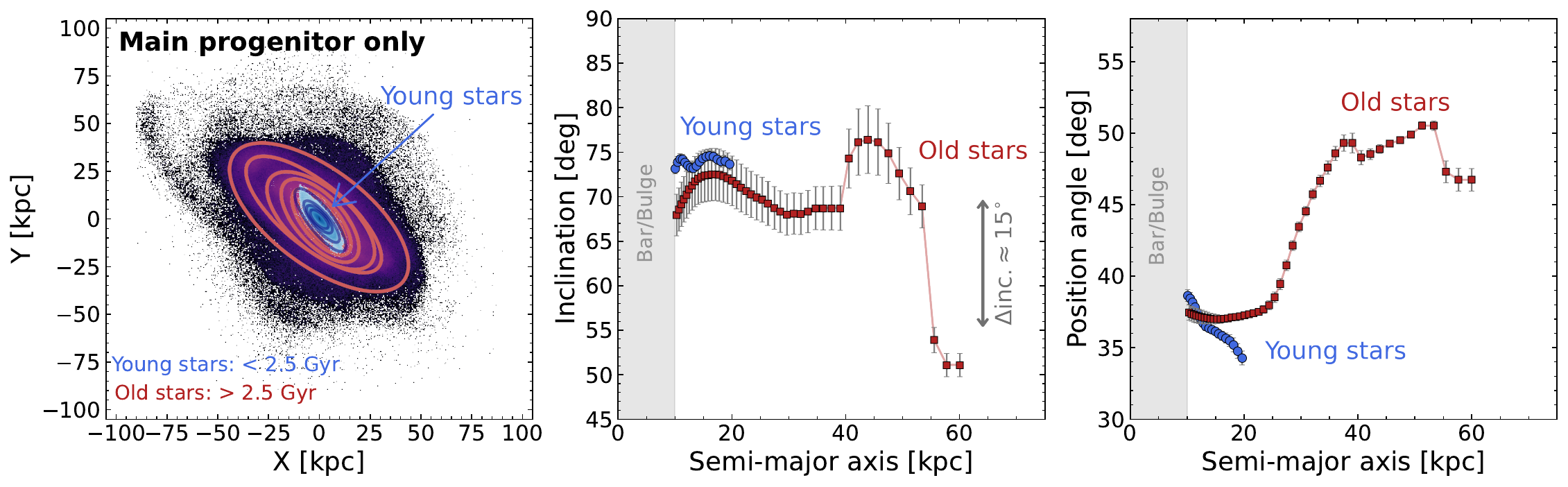}
    \caption{ Left panel: Spatial distribution of modelled stars illustrated as a 2D histogram for the older ($> 2.5$~Gyr; magma) and younger ($< 2.5$~Gyr; blue) populations. Overlaid are the best-fitting ellipses derived with \texttt{PHOTUTILS}. Blue and red ellipses correspond to the young inner disc and the older extended disc, respectively. Middle panel: Variation of the disc inclination as a function of the semi-major axis. Right panel: Position angle of the fitted ellipses versus semi-major axis. The grey shaded area in the middle and right panels indicates the central bar/bulge dominated region where ellipse fitting is less constrained.}
    \label{fig:inclination}
\end{figure*}

We focus the current analysis on the disc structure(s) that emerge following the coalescence of the two progenitors. For a detailed discussion of the Giant Stellar Stream (and the main inner halo substructures) -where tidal debris from the secondary galaxy dominates the stellar density (see the rightmost panel of Figure~\ref{fig:model_gcs})- and their 3D morphology, kinematics, and metal content, we refer the reader to \citet{Tsakonas2025A&A...699A..56T}. 

The pericentre passage of the secondary galaxy triggers intense dynamical friction, leading to rapid coalescence. While pre-merger stars undergo violent relaxation and kinematic heating (e.g., \citealt{Hopkins2009ApJ...691.1168H}), gas particles are governed by dissipative collisional dynamics \citep{Barnes1991ApJ...370L..65B}. This allows gas to shock, cool, and settle into a thin rotational plane, where a subsequent burst of star formation creates the thin young stellar disc seen in Figure~\ref{fig:model_gcs}.

The old, pre-merger stars that comprise the two merging discs are dynamically heated from the collision and attain larger velocity dispersions. A prominent tidal tail originating from stars (and gas) of the main, pre-merger M31 disc, formed after the second pericentre passage, re-accretes in the outer parts of the remnant. Thus, stellar and gaseous material falls back onto the remnant, assembling into a massive toroidal structure that encircles it, leading to the disc growing from inside out \citep{Arnaboldi2022A&A...666A.109A}. This material may be traced by old stars in the remnant outskirts and is the main stellar structure formed from the major merger that may explain numerous traits of the observed M31 disc outer edges (for the gaseous counterpart of this tidal structure and its effect on the rotation curve of M31, see \citealt{Hammer2025A&A...694A..16H}).

Figure~\ref{fig:model_gcs} illustrates the surface density of both the aforementioned structures. Stars younger than 2.5~Gyr\footnote{This timing (2.5~Gyr) is considered as the decisive age benchmark of the merger, as it roughly coincides with the star formation burst, right after the secondary passage ensues. See Appendix B  in \citet{Tsakonas2025A&A...699A..56T}, where the timing is discussed in tandem with the observed star formation history of M31.} (see blue, embedded disc confined to the inmost region) are delineated by the overlaid ellipse. The rest of the stars are older than 2.5~Gyr, representing the pre-merger stellar population of the two discs. These pre-merger, older stars are dynamically heated from the impact and rotate more slowly than the inner disc components, featuring a misaligned angular momentum relative to the younger disc. 

\subsection{The thickness of the younger and older stellar discs}

We project the simulated remnant onto the sky plane according to the inclination (77$^\circ$) and position angle (PA; 38$^\circ$) of M31 \citep{Geehan2006MNRAS.366..996G}. We employ \texttt{PHOTUTILS} \citep{photutils_larry_bradley_2025_17129028} to fit elliptical isophotes to the modelled stars, thereby delineating isodensity contours. We include only particles from the main progenitor since we aim to trace the disc structure of the remnant and not the inner halo substructure debris. Thus, particles from the secondary progenitor are excluded from the isophote fitting, as they mostly comprise inner halo stellar debris and follow high-eccentricity orbits.

The pre- and post-merger age discs have distinct kinematic and spatial properties. To quantify the spatial variation of these two populations, we calculate the final inclination and PA for the pre-merger ($>$2.5~Gyr) and post-merger ($<$2.5~Gyr) age groups independently. 
 
\texttt{PHOTUTILS} determines the inclination (defined as $q = \cos(i)$, where $q = b/a$ is the axis ratio) of each fitted isophote from the axis ratio of the corresponding ellipse. This calculation assumes that the projected stellar density traces a razor-thin, inclined disc. However, this simplification introduces a systematic bias for the simulated merger remnant, as it exhibits a remnant disc with an age-dependent thickness. This is true for M31 as well, as its disc is known to be significantly thick \citep{Dorman2015ApJ...803...24D, Dalcanton2023AJ....166...80D}. We therefore account for the intrinsic thickness of each stellar population to obtain a more accurate estimate of its inclination. To this end, we apply the standard inclination correction for an oblate spheroid of finite thickness \citep{Hubble1926ApJ....64..321H}. The corrected $i_{corr}$, is related to the observed axis ratio, $q$, and the intrinsic thickness of the disc, $q_0$, as follows:
\begin{equation}
\cos^2 (i_{corr}) = \frac{q^2 - q_0^2}{1 - q_0^2}
\end{equation}

In this formulation, a $q_0=0$ represents a perfectly thin disc. By adopting population-specific values for $q_0$, we mitigate the systematic bias that would otherwise result from a razor-thin disc approximation.

The younger (thin) and the older (thick) disc should be roughly co-planar in the inner parts of the merger remnant. We employ this configuration as a structural constraint, and we assume appropriate values for the intrinsic flattening of each disc such that \texttt{PHOTUTILS} fitting yields comparable values for their inclination and PA, in the inner $\sim$20~kpc. We adopt an intrinsic flattening of $q_{0}$$_{\_young}$ = 0.14 and $q_{0}$$_{\_old}$ = 0.31, to account for the finite thickness of the younger and older stellar discs, respectively. As a consistency check, we note that \citet{Dorman2015ApJ...803...24D} report an average dispersion of young stars in M31 that is 50\% higher than the younger component of the MW disc. For older stars, \citet{Dorman2015ApJ...803...24D} found an average dispersion (probed by red giant branch stars; RGB) that is roughly three times as high as the old stellar disc of the MW \citep{2004A&A...418..989N}. The thickness of a stellar disc is tightly linked to the velocity dispersion of the constituent stellar population. Our adopted values for the intrinsic thickness of the two age discs are consistent with the kinematically hot disc of M31 and are representative values that approximate the (varying) thickness of the two populations. We note that the direct estimation of $q_0$ from observations is complicated due to the difficulty in estimating the vertical component of the velocity dispersion ($\sigma_{z}$), as well as the scale length and scale height of M31's disc for different stellar age populations. The assumed thickness will additionally depend on certain parameters of the simulated remnant, like the selection of the stellar age groups and the assumed plane of rotation.

We aim to probe the radial variation of the geometry (PA and inclination) of the disc (up to $\sim$60~kpc) in the merger remnant. The younger disc extends only out to $\sim$20~kpc. Thus, we keep $q_{0}$$_{\_old}$ = 0.31 fixed across the entire extent of the older stellar age disc, to assess the radial variation of its PA and inclination in the outer regions. 

Figure~\ref{fig:inclination} illustrates the inclination (middle panel) and PA (rightmost panel) of the fitted ellipses after the intrinsic thickness of each population is taken into account. In the innermost regions of the remnant disc, the inclination is nearly the same for both young and old stars. 
However, from a semi-major axis distance (R$\rm_{SMA}$) of $\sim$40~kpc, the inclination of older stars diverges significantly from the inner disc structure. In particular, for large projected distances (R$\rm_{proj}$ $>$ 50~kpc), the inclination of older stars drops almost by $\sim$15$^\circ$. This indicates that the outer stellar disc becomes less inclined, appearing more face-on to the observer. This change in PA agrees with the variation ($\sim$12$^\circ$) of the inferred inclination of the inner versus outer disc stellar populations reported in \citetalias{Ibata2005ApJ...634..287I} (see also Table~\ref{tab:summary_m31_stars}). Therefore, the observed M31 disc also appears more face-on at large radii.

The young disc, confined within R$\rm_{SMA}$$\sim$20~kpc, is tilted relative to the outer disc’s stellar distribution. These pre-merger-born stars extend to much larger radii (40$-$70~kpc) in the simulated merger remnant. Beyond $45$ kpc, the fitted PA of the older stellar population exhibits a steady increase with increasing semi-major axis distances. This monotonic twist of the orbital plane of the older stars results in their projected sky positions below (in the north-east) and above (in the south-west) the photometric major axis of the inner galaxy. Notably, this spatial configuration follows the observed spatial distribution of the two prominent groups of inner halo GCs (see Section~\ref{section:GCsgroups}).

To summarize, the different spatial geometries of the younger and older stellar discs in the simulated M31 merger remnant are such that the young stellar disc generated from collisional gas resides in a thin plane within $\sim$20~kpc from the galactic center. In contrast, the old, pre-merger stars of the more massive progenitor are in a heated, thicker disc, which has a declining inclination and twists (increasing PA) from R$\rm_{SMA}$$>$30~kpc with respect to its inner region. These stellar discs in the simulated merger remnant also have different radial extents:  the older component extends out to 70~kpc. These different radial extensions are in agreement with observations of stellar populations and their chemical properties in M31 \citep{Davidge2012ApJ...751...74D, Bhattacharya2022MNRAS.517.2343B, Arnaboldi2022A&A...666A.109A}. We note that in discs whose evolution in the last billion years is dominated by secular processes like the MW and NGC 628, the thin, younger stellar disc extends to larger radii than the thicker disc (see the structural parameters for the outer MW disc in \citealt{2016ARA&A..54..529B} and the extension of the dynamically cold thin disc in NGC 628 in \citealt{10.3389/fspas.2026.1685303}). Thus, the timing of the last major accretion event in a galaxy's past determines (among others) the structural configuration of its young and older discs.

In the rest of this work, we compare the properties of the older, pre-merger stellar populations of the simulated remnant with several observed measurements of resolved old stellar population tracers in the outer disc of M31. 

In Sections~\ref{section:GCsgroups} and \ref{Section: Stellar kinematics in the extended disc} we summarise the observed properties of the resolved tracers. In Section~\ref{Section: comparison with modelled stars} we present the comparison with the merger remnant properties from corresponding regions.

\section{Observational properties of resolved tracers in M31's disc outskirts}\label{section:Observational properties}

Several independent observational studies provide the kinematic measurements and chemical properties of resolved tracers (i.e., GCs; \citetalias{Veljanoski2014MNRAS.442.2929V}, stars; \citealt{Ferguson2001ApJ...559L..13F}, \citetalias{Ibata2005ApJ...634..287I}, and planetary nebulae, PNe; \citealt{2006MNRAS.369..120M,paperVI}) in the outer disc (R$\rm_{proj}\sim$30$-$60~kpc) of M31. The properties of these tracers show the presence of a massive, extended, fairly metal-rich stellar distribution, supported by rotation. 

\subsection{Globular cluster groups in the outer disc/inner halo}\label{section:GCsgroups}
 
\begin{figure*}[htp]
    \centering
    \includegraphics[width=\textwidth]{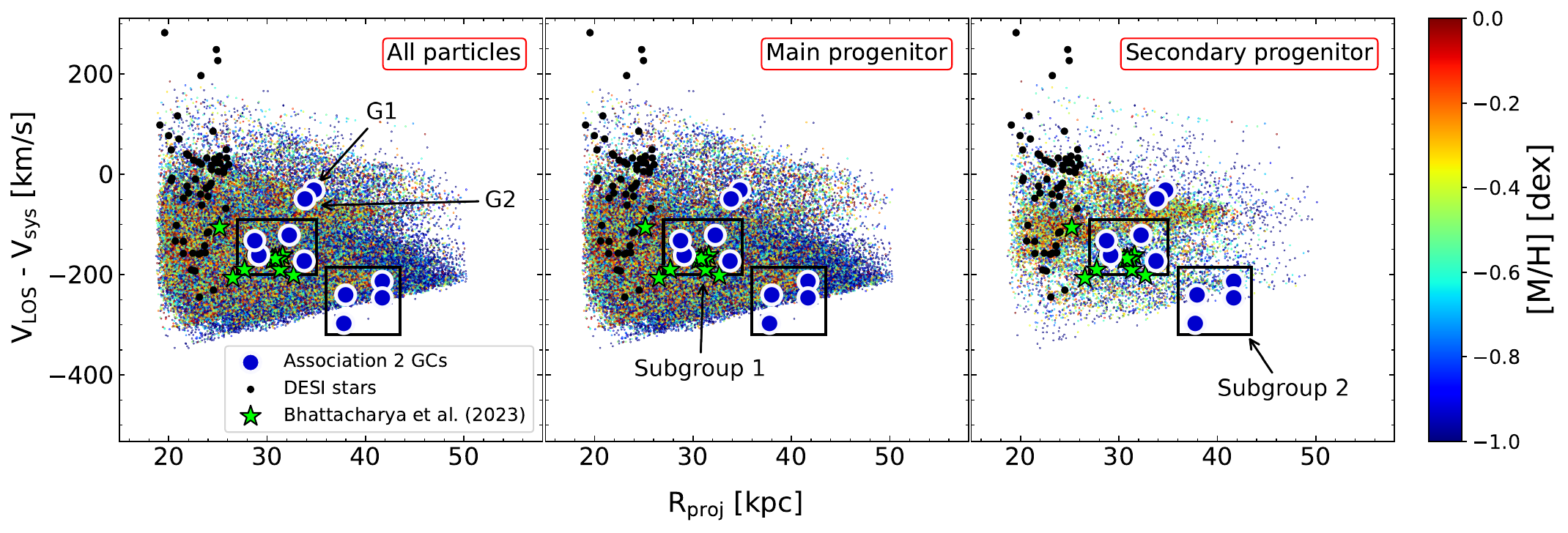}
    \caption{The phase space of modelled stars (extracted from the boxy region in the south-west, illustrated in Figure~\ref{fig:model_gcs}).  Left: Stellar particles from both progenitors. Middle panel: Stellar particles only from the main progenitor. Right panel: Stellar particles only from the secondary progenitor. Black dots in each panel represent sources in the observed DESI field close to this region \citep{Dey2023ApJ...944....1D}, while resolved PNe data from \citet{paperVI} are plotted as stars. GCs' velocities are taken from \citetalias{Veljanoski2014MNRAS.442.2929V}.}
    \label{fig:assos2_phase_space}
\end{figure*}

\begin{figure*}[htp]
    \centering
    \includegraphics[width=\textwidth]{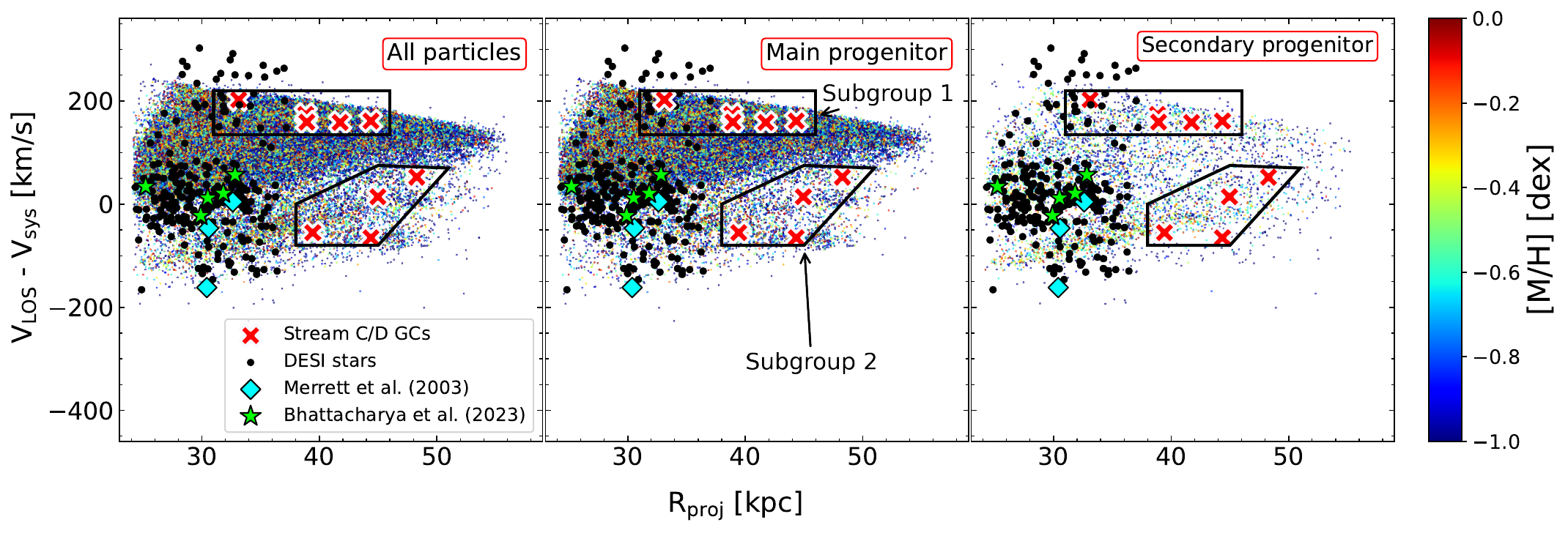}
    \caption{Same as Figure~\ref{fig:assos2_phase_space} for modelled stars extracted from the north-east portion of the remnant disc (upper left box in Figure~\ref{fig:model_gcs}). Plotted as diamonds are kinematically offset PNe in the region from \citet{2003MNRAS.346L..62M}}.
    \label{fig:streamc_d_phase_space}
\end{figure*}

The M31 GC system as a whole shows a significant rotation along the minor axis (\citetalias{Veljanoski2014MNRAS.442.2929V}). We restrict our analysis to the outer halo GC system (R$\rm_{proj}$ $>$ 25~kpc) as defined by \citet{Mackey2019MNRAS.484.1756M}. 

The inspection of the kinematics (see Figures~\ref{fig:spatial_distribution_gcs} and \ref{fig:kinematics_gcs} in Appendix~\ref{Appendix: Properties of M31 globular clusters}) of the halo GCs reported in \citetalias{Veljanoski2014MNRAS.442.2929V} reveals that the rotation signal is primarily driven by two distinct groups. These groups are located near the edges of the disc, positioned respectively above and below the photometric major axis: namely, the Association 2 GCs (hereafter Association 2) and the Stream C/D overlap GCs (henceforth Stream C/D), see Figure~\ref{fig:model_gcs}. 

Both these GC assemblies are classified as "GC-sub"\footnote{Association 2, albeit not explicitly overlapping in projection with any known stellar substructure, features two GC subgroups with notably low velocity dispersion (see Section~\ref{section: Asso2 GCs}). This coherent velocity pattern yielded a very low chance that these GCs could have formed by chance (\citetalias{Veljanoski2014MNRAS.442.2929V}); thus, Association 2 was classified as GC-sub.} in \citet{Mackey2010ApJ...717L..11M, Mackey2019MNRAS.484.1756M}. Association 2 lies along the south-west extension of the major axis (R$\rm_{proj}$$\sim$20$-$40~kpc) and dominates the blue-shifted velocity regime (see Figure~\ref{fig:kinematics_gcs} and Table~\ref{tab:ass2 parameters}). In its vicinity is a well-studied inner halo substructure, the G1 clump \citep{Ferguson2002AJ....124.1452F, Reitzel2004AJ....127.2133R, Ibata2005ApJ...634..287I, Faria2007AJ....133.1275F}. Stream C/D resides at the north-east disc counterpart (R$\rm_{proj}$$\sim$20$-$50~kpc), superimposed on a complex region where two stellar streams (Stream C and Stream D; \citealt{Ibata2007ApJ...671.1591I}) and the north-east clump \citep{Ibata2005ApJ...634..287I, Richardson2008AJ....135.1998R} intersect. Stream C/D GCs exhibit redshifted velocities, ranging from -100 to +200 $\text{km s}^{-1}$ (for the average properties of the two groups see Table~\ref{tab:summary_m31}, and for the properties of each GC Table~\ref{tab:stream CD parameters}). 

Given their primary role in driving the observed rotation within the M31 GC system, we discuss these two groups separately below. All the rotation-corrected velocities for the GC groups reported below are taken from \citetalias{Veljanoski2014MNRAS.442.2929V} and are illustrated in Figure~\ref{fig:kinematics_gcs}.

\subsubsection{Association 2 globular clusters}\label{section: Asso2 GCs}

Association 2 was initially identified and named by \citet{Mackey2010ApJ...717L..11M} as an overdensity of ten highly spatially clustered GCs at the south-west extension of M31's major axis. Through their spatial density analysis, \citet{Mackey2010ApJ...717L..11M} identified Association 2 to be the highest GC density enhancement relative to the azimuthal GC density average. This group includes G1, presumably the most massive and brightest GC of M31 \citep{Meylan2001AJ....122..830M, Ma2007MNRAS.376.1621M, Ma2009RAA.....9..641M, Nardiello2019MNRAS.485.3076N} and putatively the central star cluster of an accreted system.

Velocity measurements obtained from \citetalias{Veljanoski2014MNRAS.442.2929V} showed that these ten GCs -despite being tightly grouped in projection- do not share the same line-of-sight (LOS) velocity, see Figure~\ref{fig:assos2_phase_space}. G1 and G2 seem to be kinematically decoupled from the rest of the group, while the eight remaining clusters clearly split into two kinematic subgroups. The first subgroup consists of H7, H8, PA-22, and PA-23 with a mean rotation-corrected velocity of -63 $\pm$ 17 kms$^{-1}$ and a dispersion of 19 $\pm$ 13 kms$^{-1}$. The second subgroup contains H2, PA-18, PA-19, and PA-21 and features a mean rotation-corrected velocity of -162 $\pm$ 18~kms$^{-1}$ and a dispersion of 30 $\pm$ 28~kms$^{-1}$. The two subgroups are also spatially distinct, see Figure~\ref{fig:assos2_phase_space}; the first GC subgroup lies at an 29 $\leq$ R$\rm{}_{proj}$ $\leq$ 34~kpc, while the second group resides within an R$\rm{}_{proj}$ $\sim$38–42~kpc. The statistical analysis of \citetalias{Veljanoski2014MNRAS.442.2929V} deduced that the probability that any of the two subgroups is formed by chance is small (0.05 and 0.04 per cent, respectively). 

Using integrated light calcium-II triplet spectroscopy, \citet{Sakari2022MNRAS.512.4819S} report abundance values for two of the GCs in Association 2. They estimate a [Fe/H] = -1.34 for H7 and a [Fe/H] = -1.11 for PA-22. Both of these values are considered fairly metal-rich for halo GCs. A summary of the GCs that comprise Association 2 is given in Table~\ref{tab:summary_m31}. 

\subsubsection{Stream C/D overlap globular clusters}\label{section: stream cd GCs}

Nine GCs are found in a narrow spatial region, below the north-east extension of the photometric major axis of M31 (see Figure~\ref{fig:model_gcs}). 
\citetalias{Veljanoski2014MNRAS.442.2929V} showed that their velocity distribution features two subgroups. The first subgroup consists of H24, PA-41, PA-43, PA-45, and PA-46 (upper group in Figure~\ref{fig:streamc_d_phase_space}). It is a markedly kinematically cold group of GCs with a rotation-corrected velocity of 84 $\pm$ 4 $\rm km\,s^{-1}$ and a velocity dispersion of 8$^{+15}_{-8}$ $\rm km\,s^{-1}$. The second subgroup features B517, PA-44, PA-47, and PA-49 (lower cohort in Figure~\ref{fig:streamc_d_phase_space}) with a mean rotation-corrected 
velocity of -111 $\pm$ 49 $\rm km\,s^{-1}$ and a dispersion of 39$^{+54}_{-39}$ $\rm km\,s^{-1}$.

\citet{Sakari2022MNRAS.512.4819S} report metallicities for five out of the nine Stream C/D GCs. Three of them belong to the first subgroup (H24, PA-41, and PA-46) and the remaining two to the second (B517 and PA-44). Most of these GCs seem to be metal-poor (PA-41, PA-44, and PA-46 have [Fe/H]$\sim$-2) while H24 ([Fe/H] = -1.58) and B517 ([Fe/H] = -1.13) are found to be more metal-rich. An overview of the properties of Stream C/D GCs is given in Table~\ref{tab:summary_m31}.

\subsection{Stellar properties in the disc outskirts}\label{Section: Stellar kinematics in the extended disc}

Several studies have investigated stellar populations of the M31 disc outskirts. The observed fields are near Association 2 and Stream C/D and are discussed separately in the next Sections.

\subsubsection{The south-west region of the outer disc}

\citet{Reitzel2004AJ....127.2133R} and \citetalias{Ibata2005ApJ...634..287I} used Keck to obtain spectroscopic measurements for resolved stars near Association 2. These studies were initially motivated by the presence of the massive GC G1 (after which the G1 clump stellar overdensity is named). They both report that stars in this region are kinematically consistent with being an extension of the M31 disc. Specifically, the subtraction of the velocities of the simple exponential disc model constructed by \citetalias{Ibata2005ApJ...634..287I} from the LOS velocities of observed stars yielded residuals close to zero (see, for example, their field F1 presented in Figure~13 of \citetalias{Ibata2005ApJ...634..287I}). Thus, they concluded that stars in the remote outskirts of the M31 disc rotate fast, with a circular velocity close to that of the HI gas (v$\rm_{rot} \sim 220$ kms$^{-1}$). Their velocities are also remarkably consistent with the velocities of GCs in the first subgroup of the Association 2. 

\citetalias{Ibata2005ApJ...634..287I} report a [Fe/H] = -0.9~dex, measured from Ca$\rm_{II}$ equivalent widths, while \citet{Reitzel2004AJ....127.2133R} a mean [Fe/H] = -0.8~dex, estimated from V, I photometry. Both these values are relatively high for stars in the disc outskirts.

The G1 clump \citep{Ferguson2002AJ....124.1452F} is a well-studied stellar substructure situated near Association 2; observations in this area are critical for characterizing the local kinematic properties of the outer disc. Located at a projected distance of $\sim$29.6~kpc \citep{Faria2007AJ....133.1275F}, the G1 clump spans $\sim$12~kpc in diameter \citep{Bernard2015MNRAS.446.2789B} and has an absolute magnitude of $M_{V}$ $\simeq$ -12.6~mag \citep{Ferguson2002AJ....124.1452F}. The investigation of the properties of its stellar population has resulted in its classification as "disc-like", i.e., among stellar substructures likely formed by the heating or disruption of the disc \citep{Ferguson2005ApJ...622L.109F, Richardson2008AJ....135.1998R}. Also, \citet{Faria2007AJ....133.1275F} used HST/ACS to probe a field within the G1 clump. They report a relatively high mean photometric metallicity of [M/H] = -0.4~dex, and a strong similarity between the stars of the G1 clump and those of the M31 outer disc. These findings lent credence to their conjecture that the G1 clump is a fragment of the outer disc of M31. 

Furthermore, \citet{paperVI} reported $v_{LOS}$  for 11 PNe spatially close to the G1 clump. These PNe exhibit a narrow distribution with a peak at V$\rm_{LOS} = -469 \pm 8$ $\rm km\,s^{-1}$ and $\sigma\rm_{LOS} = 27$ $\rm km\,s^{-1}$, consistent with a relatively dynamically cold substructure with disc-like kinematics. This also agrees with the outer disc model and lag $v_{LOS}$ measurements for RGB stars in \citetalias{Ibata2005ApJ...634..287I} with a $\sigma\rm_{LOS} \sim 20 - 50$ $\rm km\,s^ {-1}$.

\begin{figure}[htp]
    \centering
    \includegraphics[width=0.5\textwidth]{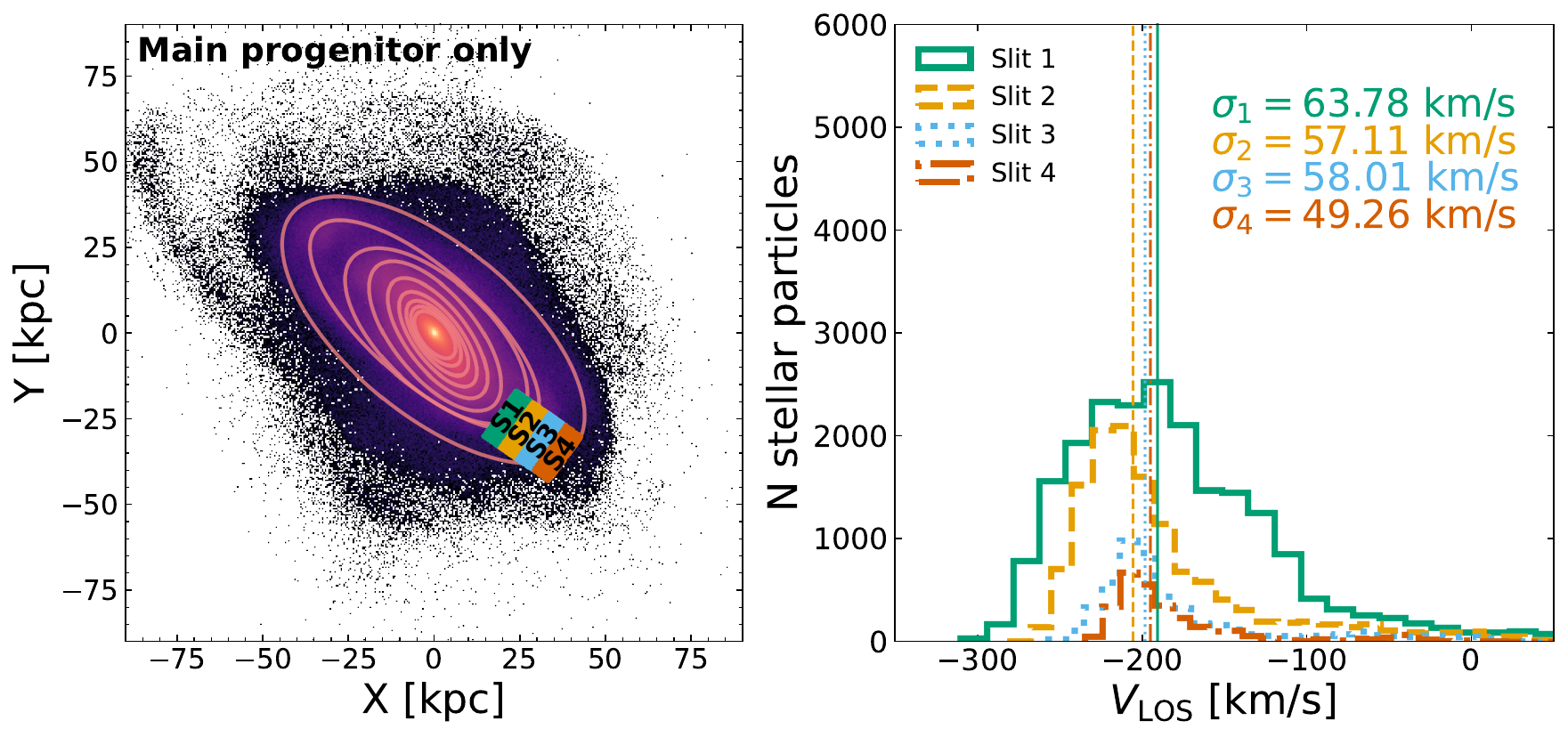}
    \caption{Left: 2D histogram of stellar particles from the main progenitor. Four slits (S1-S4) at the outer edges of the south-west portion of the modelled disc are sampled. Right: The distribution of the LOS velocities of modelled stars within each slit. The velocity dispersion (standard deviation) within each slit is shown as text.}
    \label{fig:SW_slits}
\end{figure}

\subsubsection{The north-east region of the outer disc} 

\citetalias{Ibata2005ApJ...634..287I} showed the lag $v\rm_{LOS}$ measurements for RGB stars also for a spectroscopic field near the region of the Stream C/D GCs (field F15; see their Figure~4). They report that stars in this region also display disc-like kinematics. 
On the north-east, redshifted side of the M31 disc at a major axis distance of $\sim 20$ kpc, \citet{2003MNRAS.346L..62M} detected 13 PNe with negative LOS velocity (with respect to the M31 systemic velocity). These kinematic properties are reminiscent of the lower GC subgroup of four GCs (see subgroup 2 in Figure~\ref{fig:streamc_d_phase_space}), which includes Stream C/D GCs with negative (V$_{LOS}$ - V$_{sys}$) velocities. 

\citet{Richardson2008AJ....135.1998R} presented a deep photometric field from HST reaching the outskirts of the disc. The morphology of the color-magnitude diagram is suggestive of a "disc-like" stellar population. In conjunction with its substantial rotation reported by \citetalias{Ibata2005ApJ...634..287I}, these authors also support the association to the disc for the outer fields probed by their HST pointing. Both \citet{Richardson2008AJ....135.1998R} and \citet{Bernard2015MNRAS.446.2789B} suggest that their surveyed fields near the north-east extension of the major axis and G1 clump (with strikingly similar mean RGB color and star formation histories) were formed out of disc material.

\begin{figure}[htp]
    \centering
    \includegraphics[width=0.5\textwidth]{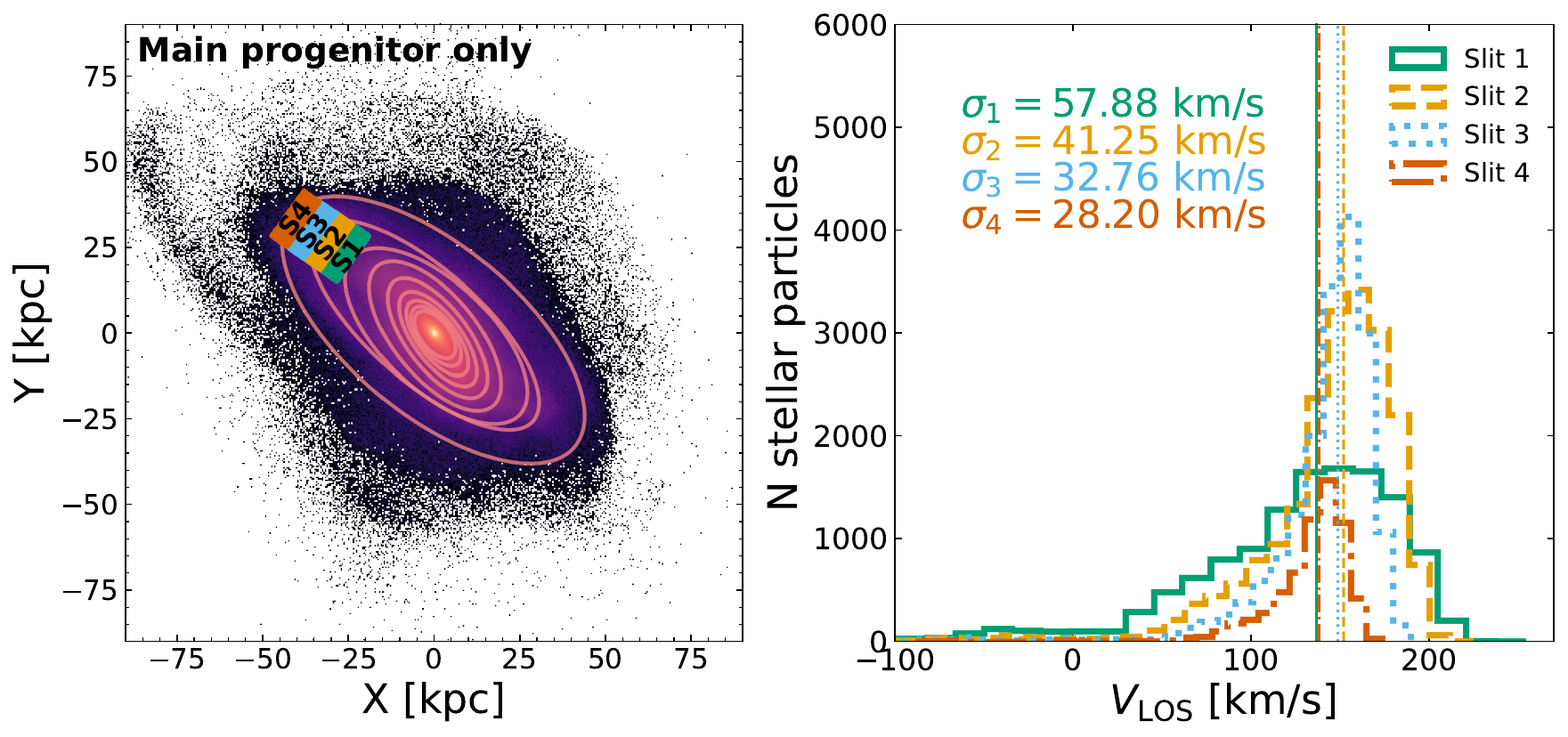}
    \caption{Same as Figure~\ref{fig:SW_slits}, with four slits sampled from the outer edges of the north-east portion of the modelled disc.}
    \label{fig:NE_slits}
\end{figure}

\section{Comparison with the simulated merger remnant}\label{Section: comparison with modelled stars}

In the major merger simulation, the spatial, kinematic, and chemical properties of each star particle are available. The outskirts of the merger remnant are dominated by old stars from the returning tidal tail, formed during the second pericentre passage. We quantify the agreement of the properties of this structure with the relevant quantities of observed discrete tracers (i.e., GCs and RGB stars).

\newcolumntype{C}{>{\centering\arraybackslash}X}

\begin{table*}
\caption{Observed average properties of the GCs of Association 2 and Stream C/D and their constituent subgroups.}
\label{tab:summary_m31}
\centering
\begin{tabularx}{\textwidth}{@{} l C C C C C C @{}} 
\hline \hline
\noalign{\smallskip}
GC groups & R$\rm_{proj}$ [kpc] & V$_{M31corr}$ [km/s] & $\sigma_{M31corr}$ [km/s] & <Age> [Gyr] & <[Fe/H]> [dex]  & <$\sigma_{[Fe/H]}$> [dex]\\
\noalign{\smallskip}
\hline
\noalign{\smallskip}
 \multicolumn{7}{c}{Subgroup 1} \\
 \noalign{\smallskip}
Association 2 & 28--34 & $-63\pm17$ & 19$\pm$13 & -- & $-1.23$ & 0.2 \\
\cline{2-7} \noalign{\smallskip}
\multicolumn{7}{c}{Subgroup 2} \\
\noalign{\smallskip}
 & 37--42 & $-162\pm18$ & 30$\pm$28 & -- & -- & -- \\
\noalign{\smallskip}
\hline
 \multicolumn{7}{c}{Subgroup 1} \\
 \noalign{\smallskip}
Stream C/D & 32--46 & $84\pm4$ & 8$^{+15}_{-8}$ & 10.54 & $-1.72$ & 0.21 \\
\cline{2-7} \noalign{\smallskip}
\multicolumn{7}{c}{Subgroup 2} \\
\noalign{\smallskip}
 & 38--50 & $-111\pm49$ & 39$^{+54}_{-39}$ & 12.61 & $-1.8$ & 0.21 \\
\noalign{\smallskip}
\hline
\noalign{\smallskip}
\hline
\end{tabularx}
\tablefoot{Observed GC velocities ($V_{M31, \rm{corr}}$) and their respective errors are taken from \citetalias{Veljanoski2014MNRAS.442.2929V}, ages from \citet{Usher2024MNRAS.528.6010U}, and metallicities from \citet{Sakari2022MNRAS.512.4819S} and \citet{Usher2024MNRAS.528.6010U} (see Table~\ref{tab:ass2 parameters} for individual GC values).}
\end{table*}

\begin{table*}
\caption{Observed average properties of stellar populations in the outer regions of M31's disc.}
\label{tab:summary_m31_stars}
\centering
\begin{tabularx}{\textwidth}{@{} l C C C C C @{}} 
\hline \hline
\noalign{\smallskip}
  & R$\rm_{proj}$ [kpc] & (i$_{inner}$ - i$_{outer}$) [$\circ$] & V$_{M31corr}$ [km/s] & <Age> [Gyr] & <[Fe/H]> [dex]    \\
\noalign{\smallskip}
\hline
\noalign{\smallskip}
 & 30--70 & 12.3 & From 170 (PNe) to 220 (HI gas) & 5.6 & -0.7 \\
\noalign{\smallskip}
\hline
\noalign{\smallskip}
\hline
\end{tabularx}
\tablefoot{The values for (i$_{inner}$ - i$_{outer}$) are obtained from the assumed inclination for the inner and the outer stellar disc in \citetalias{Ibata2005ApJ...634..287I}. The V$\rm{_{M31corr}}$ are obtained from the spectroscopic samples of PNe \citep{paperVI} and HI gas \citet{Chemin2009ApJ...705.1395C}. The age is reproduced from the average age of disc fields of \citet{Bernard2015MNRAS.446.2789B}, based on their best-fitting star formation histories. <[Fe/H]> measurements are obtained only for the south-west extension of the stellar disc. We report the mean value from the estimates of  \citet{Reitzel2004AJ....127.2133R}, \citetalias{Ibata2005ApJ...634..287I}, and \citet{Faria2007AJ....133.1275F}.}
\end{table*}

\subsection{Globular clusters groups}\label{section:Gcs compasiron with the modelled phase space}

We assess whether stars in the outskirts of the simulated disc represent a viable progenitor population for Association 2 and the Stream C/D GCs. To this end, we compare the metallicity-informed phase space distribution (R$_{\rm{proj}}$ versus the V$_{\rm{LOS}}$) of the modelled stars in two symmetric locations against the corresponding GC properties. We sample star particles in the merger remnant from two rectangular regions near the disc edges. These regions envelop Association 2 and Stream C/D GCs (without any rescaling) and are indicated by the dashed blue and red boxes in Figure~\ref{fig:model_gcs}. The phase-space properties of the stellar particles within these regions are then compared with those of the observed GC populations.

Figures~\ref{fig:assos2_phase_space} and \ref{fig:streamc_d_phase_space} show our results: for the comparison, we separately illustrate all (main + secondary progenitor; leftmost panel), main (center-most panel), and secondary progenitor (rightmost panel) particles. Thus, we may directly associate the origin of the observed tracers with the main or the secondary progenitor.

Figure~\ref{fig:assos2_phase_space} illustrates the comparison between star particles in the merger remnant and Association 2 GCs. Stellar sources from the survey fields from the  Dark Energy Spectroscopic Instrument (DESI; \citealt{Dey2023ApJ...944....1D}) in this region are also shown as full black dots. The PNe from \citet{paperVI} are illustrated as green stars. Main and secondary progenitor particles trace the two kinematically distinct GC groups. Notably, G1 and G2, albeit kinematically distinct from the rest, overlap with the edge of the wedge-like structure of secondary stars. 
Both the upper subgroup (H7, H8, PA-22, and PA-23) and the lower subgroup (H2, PA-18, PA-19, and PA-21) follow the chevrons associated with the main progenitor particles, see Figure~\ref{fig:assos2_phase_space}. 

Figure~\ref{fig:streamc_d_phase_space} presents the phase space distribution of star particles in the merger remnant within the north-east rectangular region illustrated in Figure~\ref{fig:model_gcs}. Figure~\ref{fig:streamc_d_phase_space} shows the prograde rotating population (upper subgroup; H24, PA-41, PA-43, PA-45, and PA-46) tracing the upper envelope of the wedge-like structure identified in main progenitor particles. The more kinematically spread lower subgroup overlaps with an overdensity of secondary particles. Interestingly, these stellar particles originate in a tidal tail formed after the first pericentre passage of the secondary (at a look-back time~$\sim$6-7~Gyr). This population includes GCs in retrograde motion. \citet{2003MNRAS.346L..62M} also reported the presence of PNe which are kinematically offset from the disc population in its north-east quadrant of the disc. These tracers are overplotted as cyan diamonds in Figure~\ref{fig:streamc_d_phase_space}. The examination of the properties of this GC subgroup is beyond the scope of the current work, and we refer the reader to a forthcoming publication (Tsakonas+26; in prep.). 

Through the inspection of the simulated phase space, we infer that the subgroups comprising Association 2 and Stream C/D GCs may be linked to the main or the secondary progenitor. Some GCs (lower subgroups of Association 2 and upper subgroup of Stream C/D) are traced in stars from the main progenitor. These stars are redistributed, driven by the intense external perturbation induced by the major merger, and were most probably members of the disc GCs of the pre-merger progenitor massive galaxy. Correspondingly, some GCs (G1/G2 in Association 2 and lower subgroup of Stream C/D) may have originated from the secondary progenitor.
 
\subsection{Outer disc stars}\label{Section:slits}

From the analysis of Figure~\ref{fig:assos2_phase_space} and \ref{fig:streamc_d_phase_space}, it is clear that the \citetalias{hammer} simulation predicts that most of the stars in the outermost disc are stellar particles from the main progenitor disc. We now concentrate on their kinematics, which we then compare to observations. To account for a potential radial dependence of the velocities, we select four slits along the outer edges of the north-east and south-west portions of the simulated disc, illustrated in the leftmost panels of Figures~\ref{fig:SW_slits} and \ref{fig:NE_slits}. In both regions, S1 is the slit closest to the galactic centre and S4 the farthest away from it. 

The LOS velocity distribution of stars in each slit is extracted and illustrated in the rightmost panels of Figures~\ref{fig:SW_slits} and \ref{fig:NE_slits}. Stars rotate with a velocity close to the expected velocity of disc orbits in the region, with $v/\sigma \sim$ 3-4. Such a vast extended stellar disc is consistent with the kinematics of the spectroscopic sample of \citetalias{Ibata2005ApJ...634..287I}\footnote{\citetalias{Ibata2005ApJ...634..287I} do not provide the RGB positions and individual velocities to allow a direct comparison.} (see their figure~12). In addition, the velocity dispersion steadily increases inwards (closer to the galactic centre) for both the south-west and the north-east slits, as one gets closer to the heated inner disc. 

We extract the final [M/H] of stellar particles in the merger remnant within the four slits in the north-east and their south-west counterparts using \citet{Tsakonas2025A&A...699A..56T} chemodynamical phase space. For the north-east portion of the disc we report a fairly metal-rich stellar population with median values of each slit as follows: [M/H]$\rm_{NE\_slit1}$ = -0.67, [M/H]$\rm_{NE\_slit2}$ = -0.72, [M/H]$\rm_{NE\_slit3}$ = -0.74, [M/H]$\rm_{NE\_slit4}$ = -0.80. These values are in excellent agreement with the mean metallicity estimated by \citet{Ferguson2001ApJ...559L..13F} for the outer disc (R$\sim$30~kpc) HST pointing in the north-east extension of the photometric major axis of M31 ([Fe/H]$\sim$-0.7). 

For the four slits in the south-west part of the major axis, their median [M/H] are: [M/H]$\rm_{SW\_slit1}$ = -0.59, [M/H]$\rm_{SW\_slit2}$ = -0.64, [M/H]$\rm_{SW\_slit3}$ = -0.86, [M/H]$\rm_{SW\_slit4}$ = -1.07. The standard deviation is $\sim$0.5~dex for all slits, suggestive of the complex dynamical history of the two regions \citep{Tsakonas2025A&A...699A..56T}. The metallicity values inferred from the model agree well with the metallicity measurements of \citet{Ferguson2001ApJ...559L..13F},  \citet{Reitzel2004AJ....127.2133R}, \citetalias{Ibata2005ApJ...634..287I}, \citet{Faria2007AJ....133.1275F}, and \citet{Richardson2008AJ....135.1998R}.

\section{The tidally-induced outer disc structure in the major merger: Structural and kinematic diagnostics}\label{section:Structural and kinematic diagnostics}

In this Section, we investigate the morphological and rotational properties of the post-merger disc in the major merger simulation of \citetalias{hammer}. We describe a plausible physical mechanism that gave rise to the conspicuous GC overdensities at the edges of M31's major axis and forged the vast extended disc structure around the galaxy. We employ several diagnostics to investigate the bisymmetric response of the pre-merger disc of M31 to the merger event. This dynamical response manifests itself as a precessing warp in the outer disc; indeed, as demonstrated in Section~\ref{section: Spatial configuration of the remnant}, the outer disc stellar distribution exhibits both an increased PA and a decreased inclination.  To robustly associate this structure with the major merger and evaluate its impact on the kinematic (e.g., rotational velocity, angular momentum) and spatial properties (vertical displacement, radial expansion) of disc stars, we analyze sequential snapshots from the major merger simulation of \citetalias{hammer} to trace its dynamical evolution. 

\subsection{Vertical displacement of stars}

The noticeable stellar warp emerges at large radii in the simulated remnant disc as a progressively rising vertical displacement of stars from the thin, flat plane of the inner disc. Generally, long-lived stellar warps are a well-known dynamical imprint of major mergers serving as detectable structural signatures in stellar discs (e.g., \citealt{Thulasidharan2025ApJ...993L..28T}). 

To quantify the vertical offsets of the different stellar populations from the galactic plane, we need to define a reference plane. We select two simulation snapshots. One preceding the second pericentre passage (lookback time $\sim$4~Gyr\footnote{The first pericentre passage has a moderate effect on the main disc. It only affects the less massive, secondary progenitor, generating an extended tidal tail from gravitationally extracted stars from the disc's outer regions. It is the second pericentre passage that entails the most significant perturbation on the major merger.}) and the final simulation snapshot. Defining a reference plane for the main disc (and rotating the galaxy edge-on, with respect to this plane) before the second pericentre passage is straightforward, as the disc is unperturbed and relatively thin. In contrast, the post-merger disc is kinematically hot, making the identification of a single plane challenging. We define the post-merger reference plane using stars younger than 2~Gyr; the vertical displacements are measured relative to this plane. The 2 Gyr old and younger stars are those newly formed, after the gas settles into a thin, rotationally-supported disc, with the rapid cooling giving rise to this post-merger stellar population.

\begin{figure}[htp]
    \centering
    \includegraphics[width=0.5\textwidth]{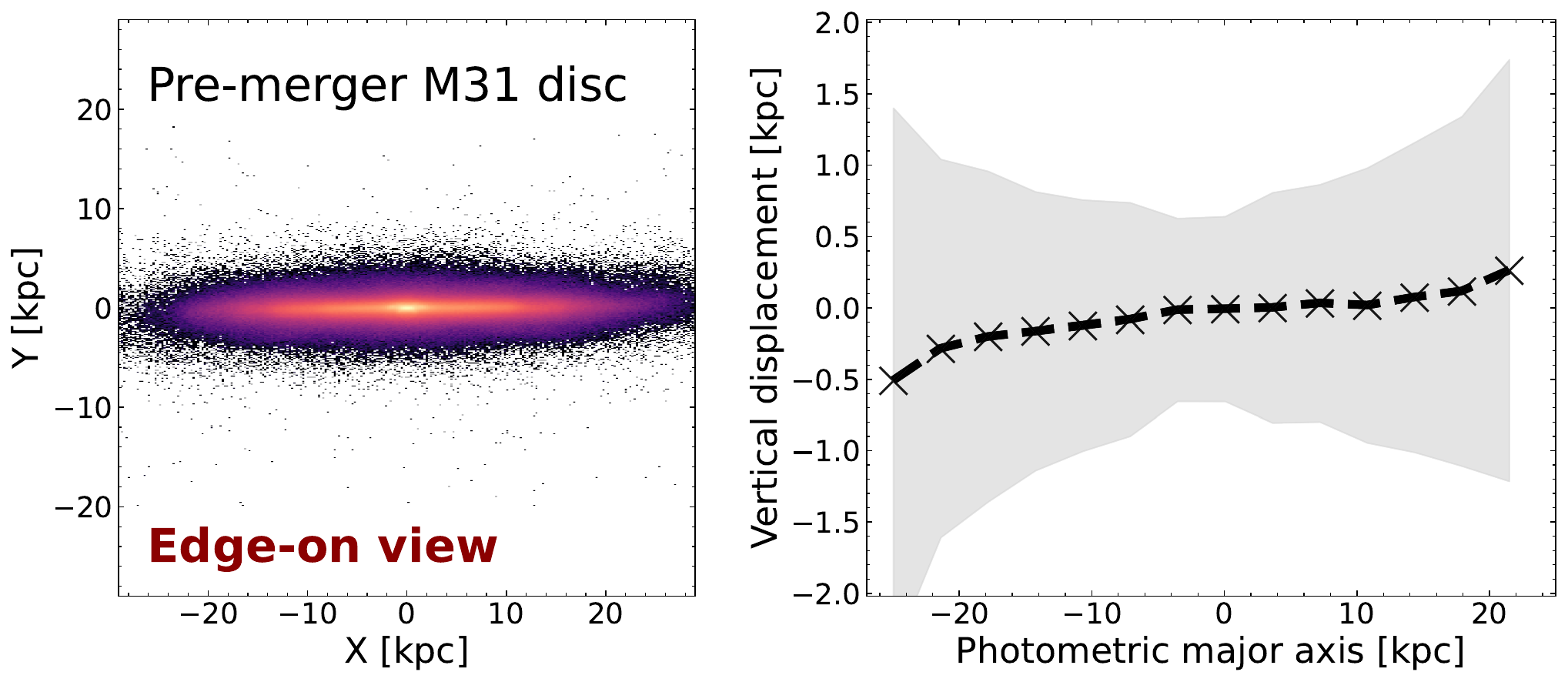}
    \caption{Left: Edge-on view of the main progenitor (pre-merger M31). Right: The median vertical displacement of main progenitor stars before the second pericentre passage.}
    \label{fig:vertical_displ_before}
\end{figure}

\begin{figure}[htp]
    \centering
    \includegraphics[width=0.5\textwidth]{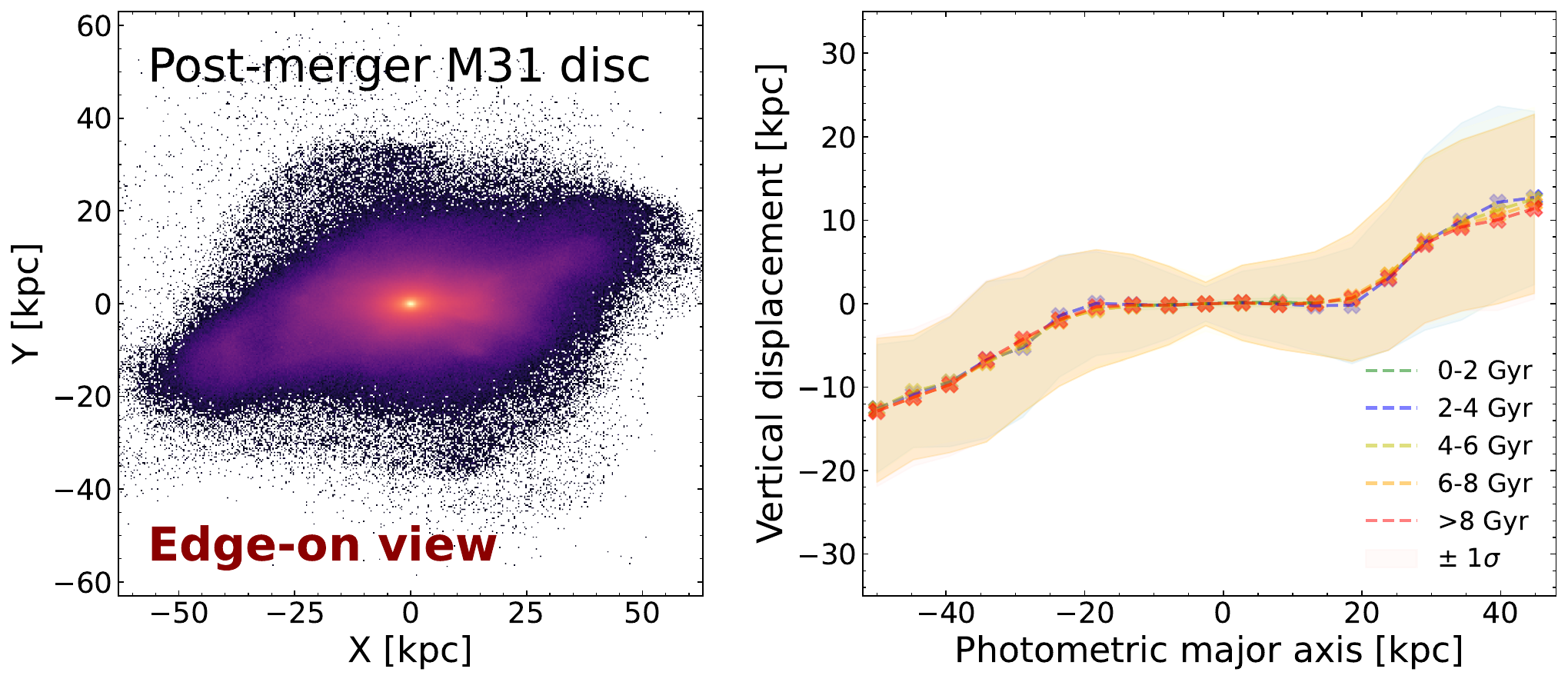}
    \caption{Left: The post-merger edge-on view of modelled stars from the main progenitor. The angles used to rotate the remnant are defined from the inclination and PA of the young (<2~Gyr old) thin disc. Right: The median vertical displacement of different stellar age groups in the remnant disc of the modelled galaxy. Stellar particles from the main progenitor only.}
    \label{fig:vertical_displ}
\end{figure}

We first assess whether the warp in the simulated analog could have formed secularly (i.e., without an external perturber). To this end, we compare the vertical structure of the main galaxy before (Figure~\ref{fig:vertical_displ_before}) and after (Figure~\ref{fig:vertical_displ}) the second pericentre passage. The right-hand panels of Figures~\ref{fig:vertical_displ_before} and \ref{fig:vertical_displ} illustrate the radial dependence (along the photometric major axis) of the off-plane displacement.  In Figure~\ref{fig:vertical_displ_before}, the pre-merger disc (i.e., only the main progenitor) is rotated in an edge-on view.  In Figure~\ref{fig:vertical_displ}, the heated, post-merger disc is also seen edge-on using the PA and the inclination of the young (<2~Gyr old) stars. There, we also define stellar age groups in 2~Gyr intervals: 2$-$4~Gyr, 4$-$6~Gyr, 6$-$8~Gyr, and $>$8~Gyr to investigate potential age-dependent variations in the vertical structure. 

This analysis reveals that the pre-merger M31 disc exhibited minimal deviation from a co-planar rotational structure. However, following the strong perturbation induced by the merger, the disc morphology changes significantly. Stars undergo substantial vertical heating, acquiring orbits that can reach far above and below the disc plane (of the order of $\sim$5$-$10~kpc). Quantitatively, the warp onsets at a semi-major axis distance of $\sim$20~kpc and peaks at the disc edges (40$-$50~kpc). 

This spatial configuration of the outer disc is directly related to the presence of the GC overdensities. Notably, both the Association 2 (Section~\ref{section: Asso2 GCs}) and the Stream C/D (Section~\ref{section: stream cd GCs}) lie at projected distances of 30$-$40~kpc from the galactic center, coinciding with the onset of the increase of the vertical displacement amplitude (warp). This extended structure exhibits disc kinematics (see Section~\ref{section:Gcs compasiron with the modelled phase space} and \ref{Section:slits}), albeit its spatial properties are misaligned with respect to the inner disc. It features a lower PA, located below the extension of the photometric major axis in the north-east and above it in the south-west, consistent with the observed spatial distribution of Stream C/D in the north-east and Association 2 in the south-west of the M31 disc.

\subsection{The post-merger radial expansion of the disc}\label{Section:radial expansion of the disc}

 To probe the merger-driven radial expansion of disc outskirts, we use modelled stars from the two dashed boxes illustrated in Figure~\ref{fig:model_gcs}. The upper box samples stars in the north-east edge of the disc, and the lower box samples stars from the south-west portion. These stars probe the regions under study by sampling stellar particles in the merger remnant from the outskirts of the remnant disc. To extract their initial galactocentric radii in the pre-merger disc and assess how the merger relocated them in the disc outskirts, we trace them at the snapshot right before the second pericentre passage. We thus map where these stars were just before the intense perturbation phase at the merger onsets.

\begin{figure}[htp]
    \centering
    \includegraphics[width=0.5\textwidth]{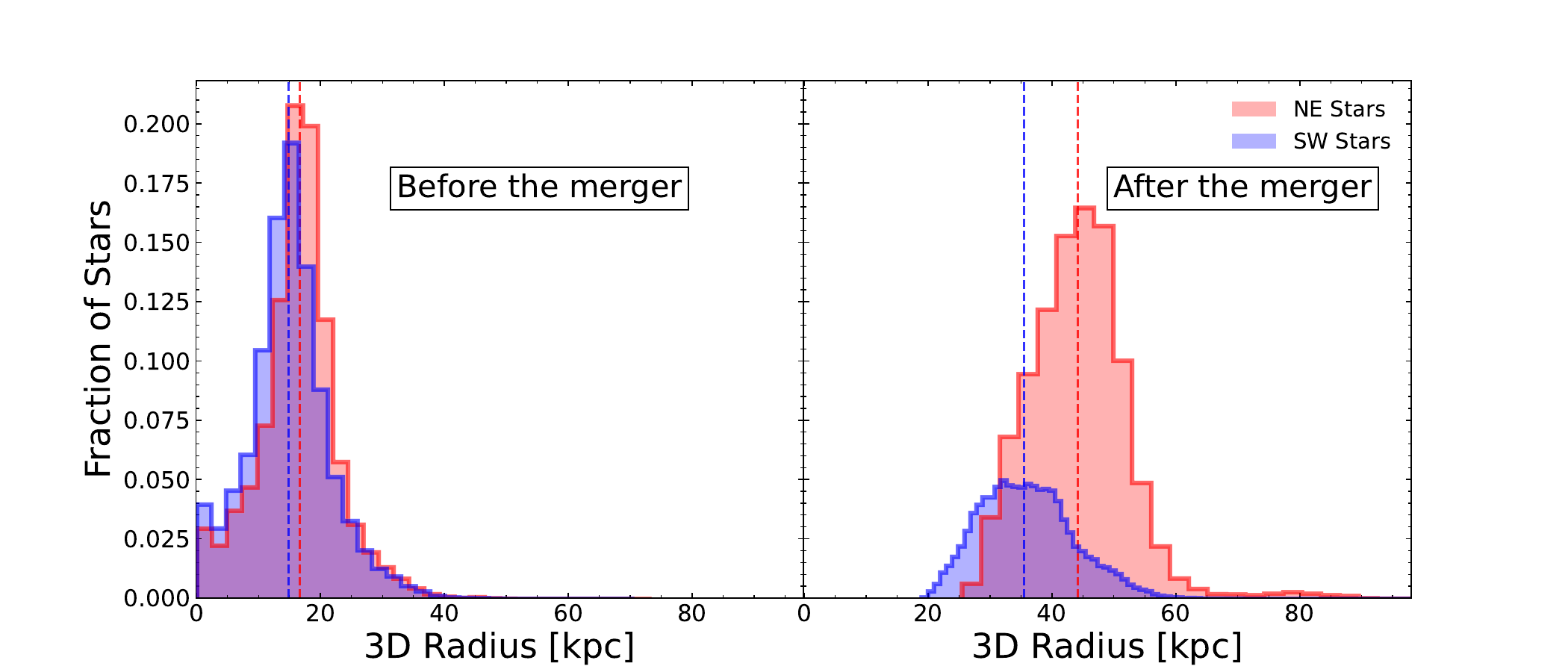}
    \caption{Spatial evolution of stars in the disc edges. The panels show the 3D distances of modelled stars selected from the north-east and south-west post-merger disc outskirts, before the second pericentre passage (left) and at the end of the simulation (right). The selected stellar particles are extracted from their final sites (the two boxes illustrated in Figure~\ref{fig:model_gcs}). They are traced back to the snapshot right before the second pericentre passage, taken at a lookback time of $\sim$4~Gyr.}
    \label{fig:gc_before_after}
\end{figure}

In Figure~\ref{fig:gc_before_after} we show that stars that are found in the outskirts of the merger remnant disc were initially placed at almost half the distance from the centre of their progenitor discs. Thus, the disc of M31 nearly doubled its size due to the major merger event. Interestingly, although the initial galactocentric distances of stars within both regions are comparable (these stars resided in roughly the same annulus within the main progenitor), stars that ended up in the north-east portion of the disc are tossed $\sim$10~kpc further. We thus deduce that the stars were subject to a stronger perturbation in the north-east than in the south-west. 

As previously discussed, we may also associate this radial expansion to the dynamical interaction after the second pericentre passage. This material is perturbed by the massive satellite and gradually re-accretes to the remnant, feeding the outer edges of the disc, after the second pericentre passage. Consequently, the major merger significantly affected M31's disc radial extent, acting as an external dynamical driver that made it nearly double in size. The GC overdensities and outer disc stellar kinematics are directly influenced by the post-merger expansion, both in terms of their spatial distribution in the merger remnant and their post-merger kinematics. The latter are discussed in the next section.

\subsection{Kinematical signature of the warp: Radial dependence of the angular momentum}\label{Section: angular momentum}

The post-merger angular momentum of disc stars is linked to the ensuing orbital configuration of the modelled stellar particles. Since the inner young disc is confined within approximately 20~kpc, effectively all stars outside this region are older than 2~Gyr, thus probing the redistributed population forged by the merger. This outer portion of the remnant disc (which underwent a merger-induced radial expansion, described in Section~\ref{Section:radial expansion of the disc}) may unfold as a coherent expanding annulus surrounding the disc (i.e., sharing smooth angular momentum radial profiles) or as a non-symmetric structure.

\begin{figure}[htp]
    \centering
    \includegraphics[width=0.5\textwidth]{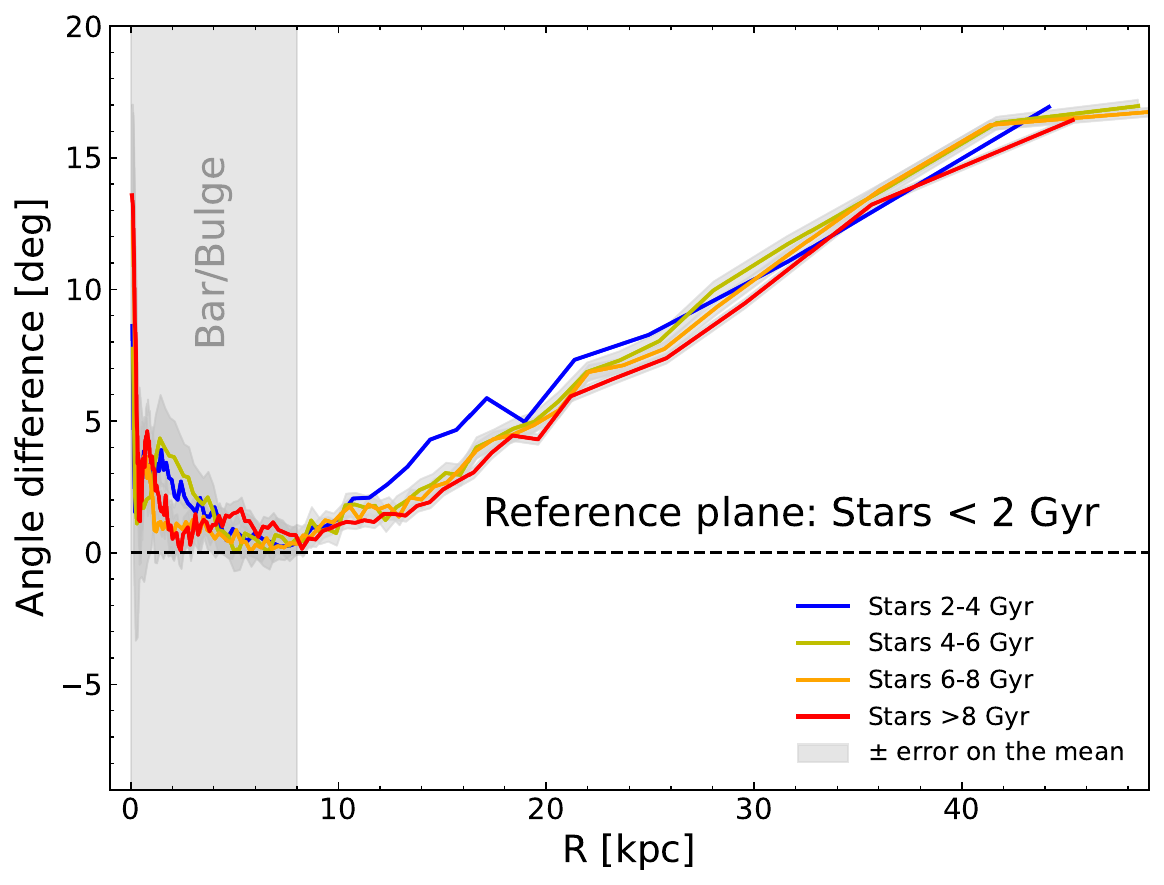}
    \caption{Radial profile of the misalignment between the angular momentum vectors of different stellar age populations. The angular offset is calculated relative to the reference plane defined by the young stellar disc ($<2$ Gyr). Profiles are shown for four distinct age bins: 2–4 Gyr (blue), 4–6 Gyr (olive), 6–8 Gyr (orange), and $> 8$ Gyr (red). Solid lines represent the median angle difference, with shaded grey regions indicating the $\pm 1\sigma$ error on the mean. The shaded region denotes the central 8~kpc, which is dominated by the bar and bulge components.}
    \label{fig:angular_mom}
\end{figure}

\begin{figure}[htp]
    \centering
    \includegraphics[width=0.5\textwidth]{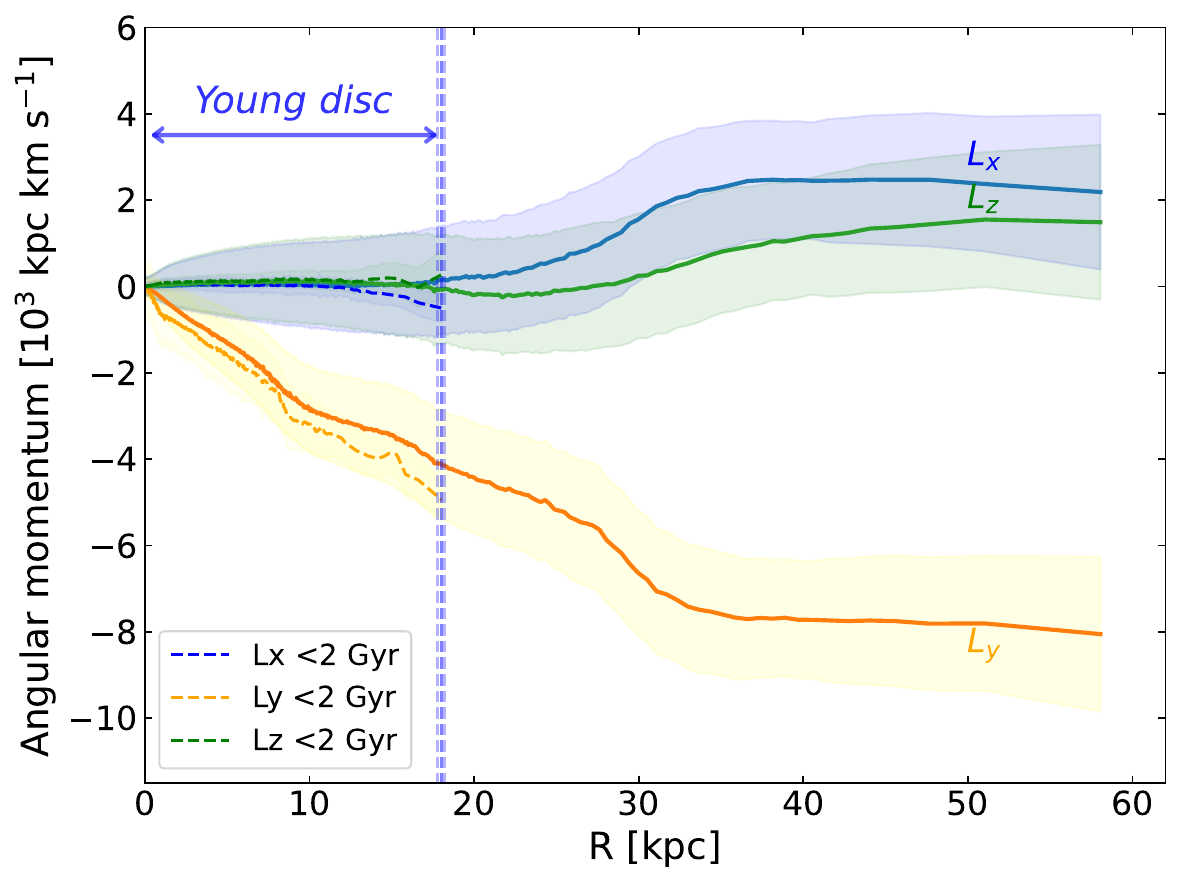}
    \caption{Radial dependence of the individual components of young (<2~Gyr; dashed lines) and old stars (>2~Gyr; solid curves reaching up to 60~kpc) in the remnant disc. The vertical dashed line specifies the region where the young disc is confined.}
    \label{fig:angu_mom_profile}
\end{figure}

We investigate the angular offset of the 3D angular momentum vector as a function of radius. To this end, we infer the angular momentum vector of stars younger than 2~Gyr, and we once again take this stellar population as the "reference plane". Once again, we select distinct age groups to assess potential age dependencies in the angular momentum properties of the stars in the disc. 

Each line in Figure~\ref{fig:angular_mom} illustrates the angle between two vectors. The angular momentum vector of young stars (which is the reference anchor) and the angular momentum vector of one of the four age groups. If each angular momentum were aligned with the respective reference plane, the corresponding angle would be null. However, we deduce that for radii larger than 20~kpc, the angular momentum vector of older stars is tilted compared to that of the young stars. Notably, this behaviour is age-independent.

Consequently, we may infer that the only kinematically distinct stellar population is the one that spawns from the post-merger, reformed gas disc ($<$2~Gyr). The rest of the disc stars behave roughly alike in terms of their angular momenta; they systematically deviate from a thin rotational plane. Severely perturbed by the interaction, their angular momentum profile becomes progressively more complex with radius, but this modulation is roughly age-independent; it appears similar for all the pre-merger stars. This behavior is reminiscent of that mapped as "re-orientation" of discs in response to massive mergers in MW-like galaxies (recently investigated by \citealt{2026OJAp....961450B} in the TNG50 cosmological simulations) and the gas-stellar kinematic misalignments studied through the CIELO simulations \citep{Casanueva2026arXiv260607160C}.

To further examine the geometry of the tidally redistributed stellar material, we assess the 3D properties of the disc angular momentum. We decompose the angular momentum of stars in the post-merger disc into its individual components (L$_{x}$, L$_{y}$, L$_{z}$) and we assess their radial evolution. Figure~\ref{fig:angu_mom_profile} illustrates the angular momentum profile of young ($<$2~Gyr; dashed lines truncated at $\sim$20~kpc) and older ($>$2~Gyr; solid lines reaching out to $\sim$60~kpc) stars from the main progenitor. We note that the axis perpendicular to the orbital plane is the y-axis. Thus, it is the L$_{y}$ component, which charts the orbit in the disc, and naturally dominates, especially in the inner regions. As expected, from a thin rotating disc, the other two components of the angular momentum (L$_{x}$
and L$_{z}$) have minimal contributions to the young stellar population. 

Collectively, we may clearly infer that in the inner regions of the disc, the angular momentum of the young stars aligns with the rest of the stars, with the L$_{y}$ component monotonically increasing. However, as the number of young stars decreases with increasing radius, the angular momentum vector of older stars acquires a tilt. There, we see that the other two angular momentum components become non-negligible and the rotation of the stars older than 2~Gyr is no longer confined to a thin co-planar disc at R$\rm_{proj}$~$>$~25~kpc (see Figure~\ref{fig:angular_mom}). 

\subsection{The temporal evolution of the tidal tail that co-forms the remnant disc}

\begin{figure}[htp]
    \centering
    \includegraphics[width=0.5\textwidth]{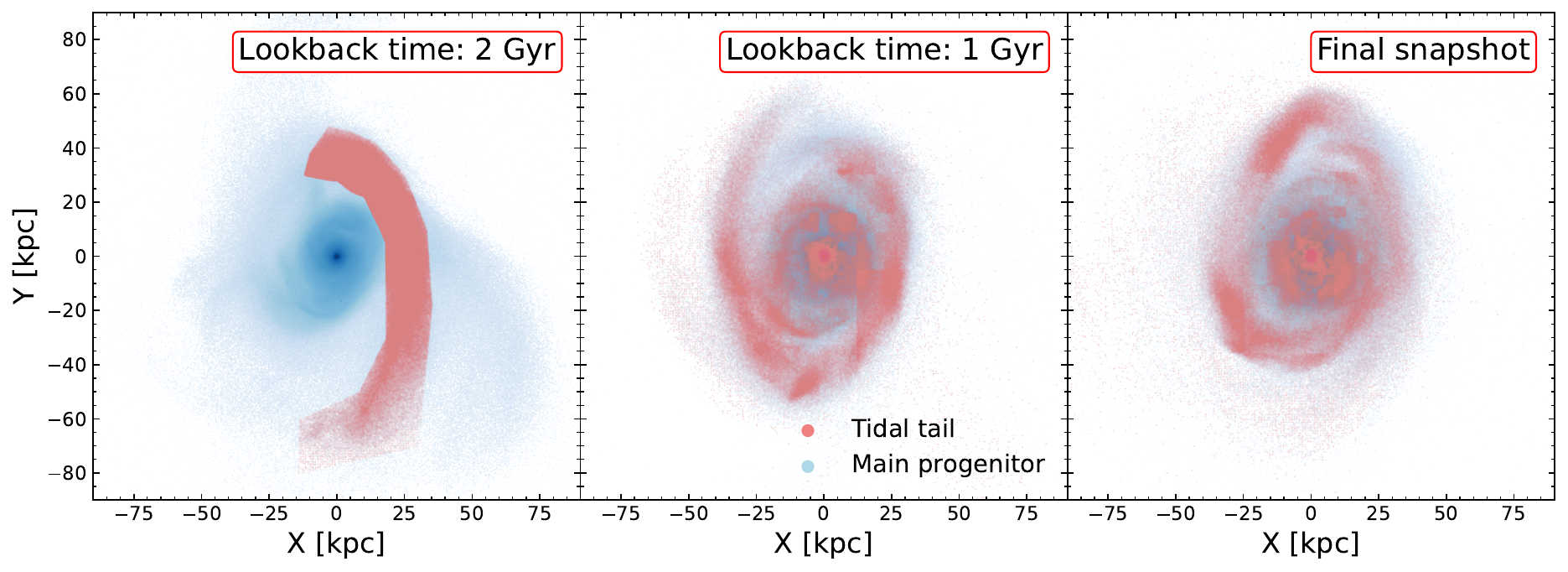}
    \caption{Time evolution of the merger-induced tidal tail at the second pericentre passage (left), at a lookback time of 1~Gyr (center), and at the final (present-day) snapshot of the \citetalias{hammer} simulation. In each panel, only main progenitor stellar particles are illustrated.}
    \label{fig:tidal tail}
\end{figure}

The \citetalias{hammer} merger model features distinct tidal tails formed during the merger. The portion of this material (stars and gas) that returns to the merger remnant causes non-axisymmetric perturbations in the outer regions. Some of these returning particles form loops warping around the merger remnant, contributing to the formation of stellar substructures, like the Giant Stellar Stream \citep{Tsakonas2025A&A...699A..56T, Hammer2025A&A...694A..16H}. 

Another tidal tail formed after the second pericentre passage from baryonic material of the main progenitor contributes significantly to the formation of the outer stellar disc of the final remnant investigated in the present paper. We spatially select this tidal tail at the time of its formation, at a lookback time of 2~Gyr (Figure~\ref{fig:tidal tail}; leftmost panel). We trace the stellar particles to a snapshot 1~Gyr later (Figure~\ref{fig:tidal tail}; center-most panel) and in the final snapshot (Figure~\ref{fig:tidal tail}; rightmost panel). This tidal tail is not completely disrupted even 2~Gyr after its formation. We also deduce that these stellar particles must have retained a significant portion of their angular momentum. Naturally, their relatively kinematically cold orbits contribute to the inside-out formation of the stellar disc. This tidal tail, formed at the time of the strongest perturbation of the merger, is re-accreting onto the merger remnant with a slightly tilted angular momentum with respect to the disc. It is a key feature of the \citetalias{hammer} simulation that drives the formation of the vast extended stellar disc around the M31 galaxy model and shapes the geometry of the outermost parts in the merger remnant.

\section{Summary and Conclusions}\label{section: summary and conclusions}

From both theoretical grounds (e.g., \citealt{Stewart2008ApJ...683..597S}) and observational studies (e.g., \citealt{Hammer2009, Rodrigues2017MNRAS.465.1157R}), mergers drive galaxy evolution and shape the observed properties of MW-mass galaxies. When the merging galaxies contain a significant amount of cold gas, the ensuing remnant morphology may readily produce an extended stellar disc shaped by merger-induced tidal forces and a spatially and/or kinematically misaligned young stellar disc \citep{Hopkins2009ApJ...691.1168H, Casanueva2026arXiv260607160C}.

M31 hosts a merger-transformed dual-disc structure. A co-planar thin and thick disc with different scale heights \citep{paperII, Dalcanton2023AJ....166...80D} and chemical abundances \citep{Arnaboldi2022A&A...666A.109A}. Intriguingly, the remote outskirts (R$\rm_{proj}$ > 30~kpc) of the merger remnant feature disc kinematics, albeit with a distorted morphology (\citetalias{Ibata2005ApJ...634..287I}). In addition, prominent GC groups cluster tightly in two symmetrical regions above (in the south-west) and below (in the north-east) the projected extension of the photometric major axis, exhibiting a notably high rotational signal (\citetalias{Veljanoski2014MNRAS.442.2929V}). However, their angular momentum vector is misaligned with respect to the inner cohort of GCs \citep{Mackey2019Natur.574...69M}. In addition, the metallicity of stars \citep{Reitzel2004AJ....127.2133R, Ibata2005ApJ...634..287I,  Faria2007AJ....133.1275F} and that of some GCs \citep{Sakari2022MNRAS.512.4819S} in these remote regions of the disc is relatively high for their galactocentric distance. 

We employ the well-motivated gas-rich major merger model of \citetalias{hammer} for the giant Andromeda galaxy to compare the spatial and kinematic properties of the merger remnant with the observational counterparts of M31's outskirts from resolved tracers. Our results may be summarised as follows:
 
   \begin{itemize}
      \item In the final product of the merger event, the (1:4) mass merger simulation by  \citetalias{hammer} predicts that the distribution of the older stars in the pre-merger progenitor disc is stretched, distorted, and warped in an extended disc structure. This is in agreement with the results by \citet{2006MNRAS.369..120M}, who found that there was no evidence for a cut-off in its exponential disc population even beyond four disc scale lengths, and no signs of a halo population that is distinct from the bulge out to ten effective bulge radii. 
     \item The merger remnant contains stellar discs with distinct geometry and formation pathways. In addition to the older, hotter, thicker, extended, and warped disc,  a younger disc is rebuilt from cold gas that retained a significant portion of its angular momentum and experienced a post-merger star formation burst. The young, thinner disc is less extended ($\sim$20~kpc) and kinematically colder than the older, hotter, and thicker disc. The two discs feature almost similar angular momentum vectors, as long as the young disc persists. Older stars, however, show a $\sim$15$^\circ$ drop in inclination (becoming more face-on) and a twist in their PA of $\sim$10$^\circ$, both ensuing for R$\rm_{proj}$ > 30~kpc. The varying inclination quantitatively agrees with the kinematically inferred spatial configuration of resolved stars in the disc outskirts (\citetalias{Ibata2005ApJ...634..287I}), while the PA twist accounts for the spatial distribution of the most prominent GC clumps at the edges of the M31 disc \citep{Mackey2019MNRAS.484.1756M}.
      \item The spatial properties and the kinematics of the prominent groups of GCs in the outskirts of M31's disc may be linked to the tidally redistributed material from the main progenitor disc of M31. Thus, they should predominantly feature a GC population originating in the massive pre-merger M31 galaxy, and its present-day properties are shaped by the last major merger of the galaxy.
      \item We investigate the kinematic properties of the extended disc structure, sampling stars from the outer edges of the disc remnant. Pre-merger stars are rotationally supported and align with the inferred kinematics of resolved tracers (\citetalias{Ibata2005ApJ...634..287I}; \citealt{paperVI}).
      \item We compared the vertical displacement of pre- and post-merger stars of the main progenitor galaxy. We report a merger-driven vertical heating for post-merger stars, consistent across all age groups and of the order of $\sim$5$-$10~kpc (see Figure~\ref{fig:vertical_displ}, right panel). 
      \item The post-merger disc of M31 underwent significant radial expansion. We sample modelled stars from two rectangular boxes residing in the remnant outskirts in a snapshot before the second pericentre passage. We report an asymmetric expansion of the post-merger disc, with its north-east portion featuring stars spread out over larger distances. 
      \item The angular momenta of the young and older stars are aligned in the inner disc out to $\sim$20-30~kpc, where these two structures are nearly co-planar. At larger radii where mostly old, pre-merger stars are found, their angular momenta are distinct from that of the inner disc. Specifically, the radial angular momentum profile of the post-merger stellar population showed that the outer disc warped structure is a complex, twisted, and lopsided structure, and not a symmetric annulus undergoing a post-merger radial expansion.
     \end{itemize}

We have presented a comprehensive picture of the vast stellar disc in M31, demonstrating how its present-day properties emerged from the massive merger event in its recent 2-4 Gyr past. Utilizing the major merger framework established by \citetalias{hammer}, we showed that this event serves as the primary dynamical determinant for the galaxy's vast, extended stellar disc. Following the second pericentre passage of the secondary galaxy and the subsequent onset of the coalescence, tidal material from the more massive progenitor is gravitationally redistributed. This material torques and warps around the remnant, effectively feeding the disc from the outside. This pre-merger stellar population provides a robust explanation for the origin of M31’s outer disc, a structure that has long remained elusive \citep{Ferguson2001ApJ...559L..13F, Ibata2005ApJ...634..287I, 2006MNRAS.369..120M}.

Indeed, M31 occupies a unique position in extragalactic studies. It is currently the only massive spiral galaxy for which a tailored, high-resolution analytic model of its accretion history is available alongside resolved stellar chemodynamical data. This synergy renders M31 a valuable laboratory for near-field cosmology. By investigating the merger-induced warp in the outer disc, we establish a definitive link between a galaxy's merger history and its distinct morphological features. Our findings provide a cohesive framework that connects the spatial, chemodynamical, and structural properties of M31 to its last major accretion event - a typical process within $\Lambda$CDM cosmology.

Our results are also timely since instruments such as the Prime Focus Spectrograph on the Subaru Telescope \citep{TamuraSubarupfs} will obtain resolved spectra for Local Group stellar populations, including M31 \citep{Chiba2026arXiv260409875C}. The imminent launch of the Roman Space Telescope (with a unique combination of wide-field, high spatial resolution, and low background) offers, for the first time, the opportunity to acquire a full 3D space motion for stars in its inner halo and robust space motion for its entire GC system \citep{Dey2023arXiv230612302D}. Roman is capable of achieving photometric depth deeper than the PAndAS survey (median, 5$\sigma$ g and i depths of 26.0 and 24.8, respectively; \citealt{Ibata2014ApJ...780..128I}), which is necessary to probe the stellar population overlapping with Stream C/D GCs. 

Furthermore, the next generation of wide-field facilities is poised to reveal a plethora of merger-inflicted disc galaxies across the local Universe. Upcoming large-scale surveys -including those from the Vera C. Rubin Observatory/LSST and Euclid- will provide the wide-field imaging necessary to identify similar tidal signatures and warped morphologies in more distant systems. In this context, M31 serves as the foundational archetype. The detailed study and juxtaposition of the M31 remnant against simulated analogues provides a critical guideline for interpreting the complex datasets and modeling efforts of the coming decade.

\begin{acknowledgements}
CT wants to thank the European Southern Observatory (ESO) for the opportunity to visit ESO, Garching, Germany, under the Early-Career Scientific Visitor scheme. CT thanks Andreas Burkert and the Max-Planck-Institut für extraterrestrische Physik for support for his visit in 2025.  CT would like to thank Claudia Pulsoni, Lucas M. Valenzuela, Marina Rejkuba, Lodovico Coccato, and Eric Emsellem for fruitful discussions on the dynamics of GCs. CT thanks Michele Bellazzini for the discussion on the RBC.

\end{acknowledgements}

\section*{Software}
This research output made use of Astropy, a community-developed core \texttt{PYTHON} package for Astronomy \citep{python2007CSE.....9c..10O}, \texttt{PHOTUTILS}, an Astropy package for detection and photometry of astronomical sources \citep{photutils_larry_bradley_2025_17129028}, as well as \texttt{NUMPY}, \texttt{SCIPY}, and \texttt{MATPLOTLIB} \citep{harris2020array, 2020SciPy-NMeth, Matplotlib2007CSE.....9...90H}. We also made use of NASA’s Astrophysics Data System Bibliographic Services\footnote{https://ui.adsabs.harvard.edu}.
\bibliographystyle{aa} 
\bibliography{ref.bib}

%
%

\begin{appendix} \label{appendix}

\section{Properties of M31 globular clusters}\label{Appendix: Properties of M31 globular clusters}

We summarize some basic properties of M31's GC system for reference. Particular emphasis is given to the properties relevant to the predictions of the major merger and the scientific case under discussion.

\subsection{Spatial distribution}

\begin{figure}[htp]
    \centering
    \includegraphics[width=0.5\textwidth]{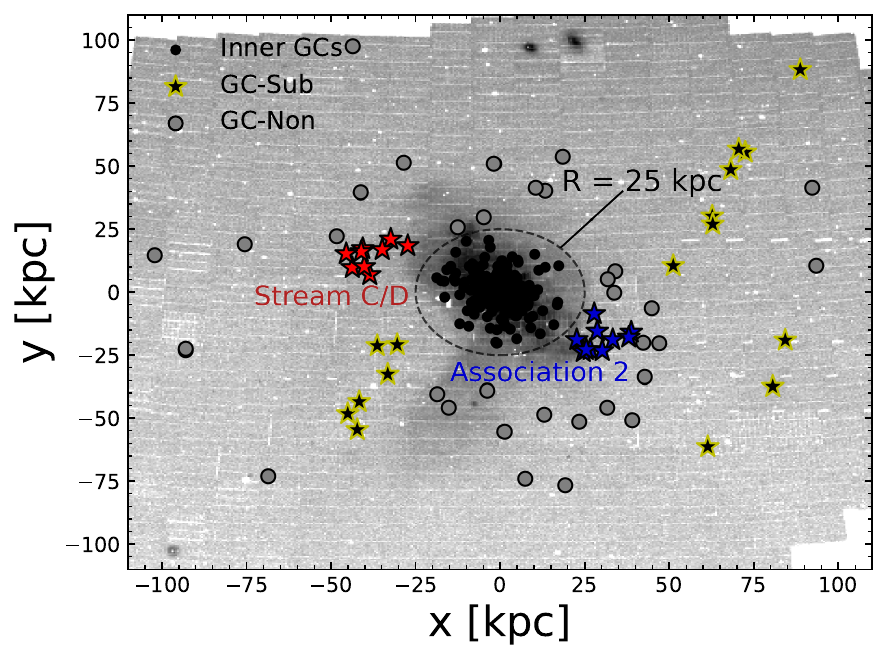}
    \caption{Spatial distribution of GCs on top of the PAndAS number density map. GC coordinates are obtained from \citet{Caldwell2016ApJ...824...42C}. The dashed circular region distinguishes inner (R$\rm_{proj}$ $<$ 25~kpc) and outer (R$\rm_{proj}$ $>$ 25~kpc) GCs.}
    \label{fig:spatial_distribution_gcs}
\end{figure}

\citet{Caldwell2016ApJ...824...42C} published a complete census of high signal-to-noise M31 GC spectra obtained with MMT/Hectospec \citep{Fabricant2005PASP..117.1411F}. Assuming a cut-off radius of R$\rm_{proj}$ = 25~kpc, we may split M31's GC population into inner GCs  (R$\rm_{proj}$ $<$ 25~kpc) and outer (R$\rm_{proj}$ $>$ 25~kpc) GCs. Figure~\ref{fig:spatial_distribution_gcs} illustrates the spatial distribution of all confirmed M31 GCs. The majority ($\sim$80\%) of M31 GCs are confined within the premises of its stellar disc, within 25~kpc.  

\citet{Mackey2019MNRAS.484.1756M} used the complete sample of outer halo M31 GCs to quantify their physical association with underlying stellar substructures. They calculated the average density of metal-poor giants in the halo of M31 from the PAndAS resolved stellar map \citep{McConnachie2018ApJ...868...55M}, and they reported that a fraction of GCs reside preferentially on top of higher-than-average stellar density regions. Based on this spatial overlap, they categorize outer halo GCs into three subgroups: i) substructure GCs (GC-sub), ii) non-substructure GCs (GC-non), and iii) ambiguous GCs (GC-amb).

To ensure consistency with previous studies, we adopt the GC classification scheme (GC-sub, GC-non, and GC-amb) and the names of the GC groups denoted in \citetalias{Veljanoski2014MNRAS.442.2929V}. These subgroups include Association 2, Stream C/D, Stream C, Stream D, the South-West Cloud, and the North-West Stream. For visual clarity, in all subsequent figures, GC-sub members are denoted by star symbols, while circles represent GC-non members. We note that the spatial coincidence with tidal debris or the spatial clustering of several GCs does not necessarily imply a physical association. Indeed, as noted by \citetalias{Veljanoski2014MNRAS.442.2929V}, several of these groups do not constitute a single kinematically coherent ensemble.

\subsection{Kinematics} \label{section: Kinematics}

To assess the chemodynamical properties of M31 GCs, \citet{Caldwell2016ApJ...824...42C} divided their sample into three metallicity bins: metal-rich (-0.4 < [Fe/H]), intermediate (-0.4 < [Fe/H] < -1.5), and metal-poor ([Fe/H] < -1.5) GCs. They demonstrated that the GC metallicity distribution is not bimodal (unlike that of the Milky Way, which shows two distinct metallicity peaks; \citealt{Massari2019A&A...630L...4M, Leaman2013MNRAS.436..122L}). These metallicity intervals reveal a clear kinematic segregation: the metal-rich group traces the rotation of the disc, while the other two groups show milder prograde rotation with lower amplitude and higher dispersion. 

\begin{figure}[htp]
    \centering
    \includegraphics[width=0.5\textwidth]{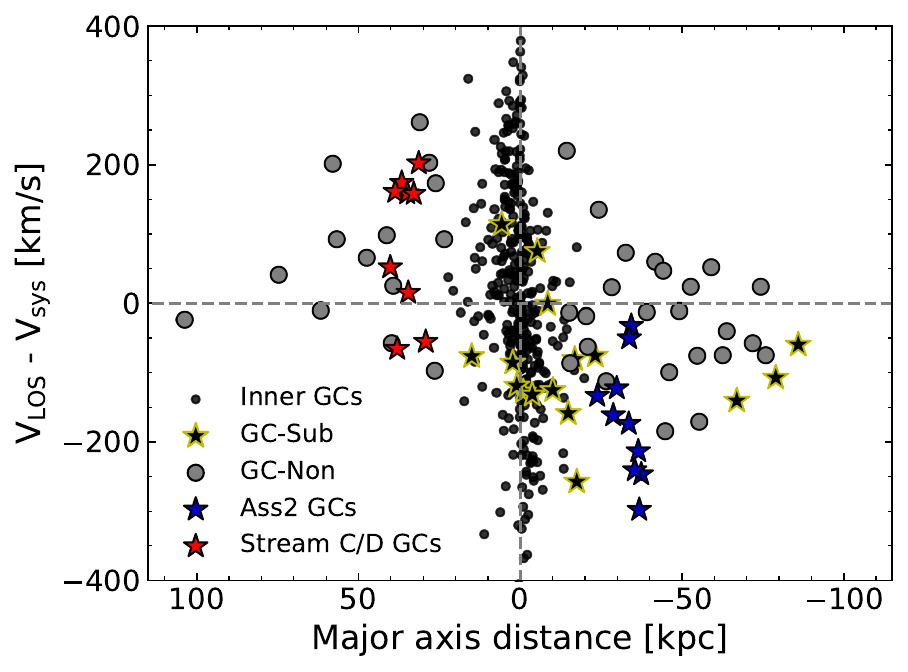}
    \caption{GCs LOS velocities corrected for M31 systemic velocity versus projected distance along the major axis. The two groups that underpin this study (Association 2, Stream C/D) are plotted with different colors.}
    \label{fig:kinematics_gcs}
\end{figure}

\citetalias{Veljanoski2014MNRAS.442.2929V} reported radial velocity measurements for M31 GCs residing in the outer halo. Their analysis demonstrated a significant degree of net rotation exhibited by this population, with the same rotation axis and direction as the inner GCs, albeit with a lower amplitude. Both the inner (black dots) and the outer group velocities (both M31 corrected) are plotted with respect to their projected major axis distance in Figure~\ref{fig:kinematics_gcs}.

\section{Ages and metallicities of M31 globular clusters}\label{gradient appendix}

\citet{Usher2024MNRAS.528.6010U} used spectroscopically informed priors to estimate ages and metallicities for a subsample of 290 M31 GCs.  In contrast to the MW's GC population, which is almost exclusively older than 7~Gyr \citep{Massari2019A&A...630L...4M}, M31 hosts a significant number of younger clusters. \citet{Usher2024MNRAS.528.6010U} reported a broad age distribution that includes $\sim$20 clusters younger than 5~Gyr. Given the recent major merger and the associated disc starburst, determining the spatial properties of these different age groups is relevant. Specifically, analyzing their spatial distribution will allow us to link distinct stellar populations in the simulation (e.g., stars formed before versus after the merger) to observed GCs with similar properties.

\begin{figure}[htp]
    \centering
    \includegraphics[width=0.5\textwidth]{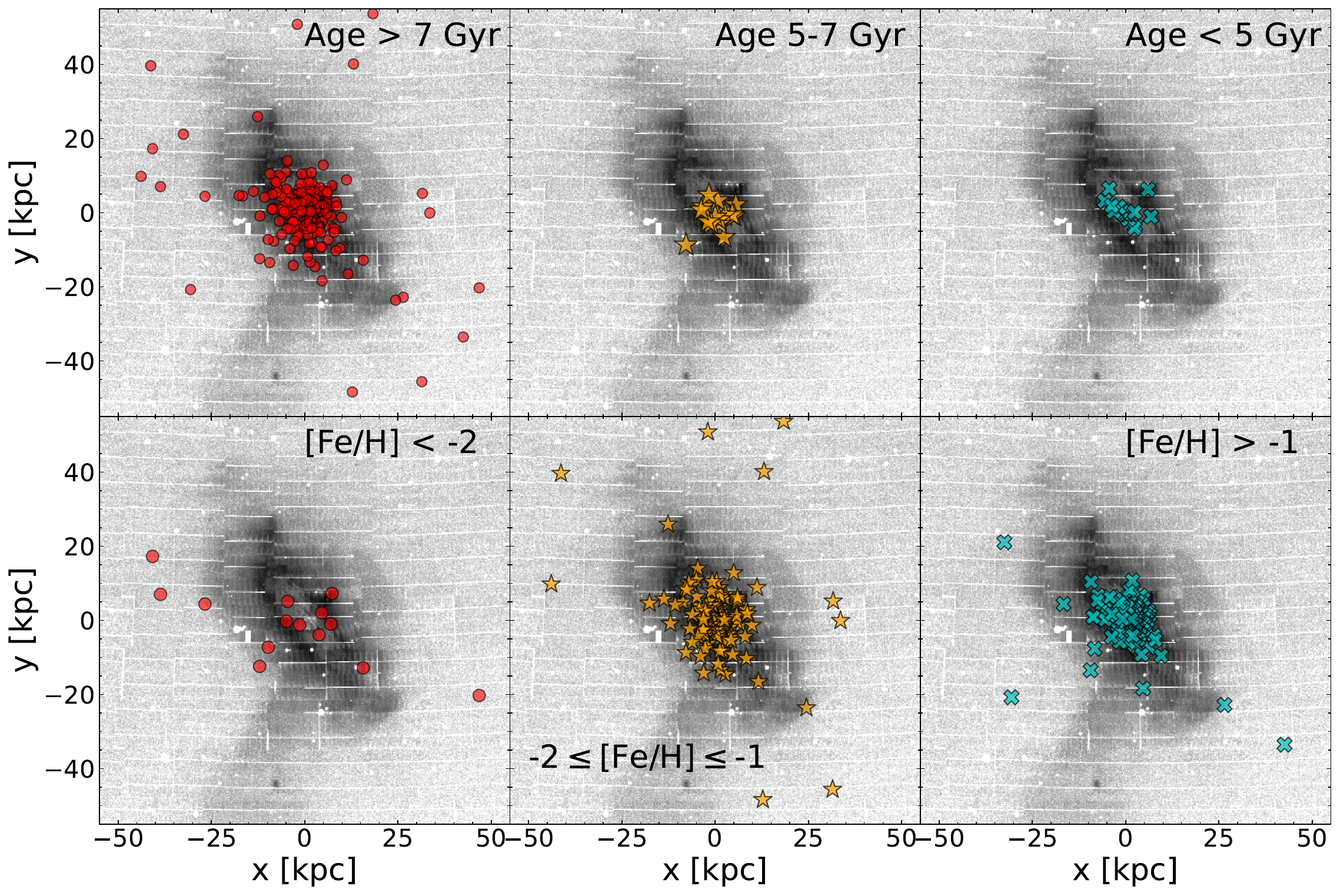}
    \caption{Upper row: The three age bins of M31 GCs (as reported in \citealt{Usher2024MNRAS.528.6010U}) overlaid on the PAndAS number density map. Lower row: The three metallicity bins of M31 GCs (as reported in \citealt{Usher2024MNRAS.528.6010U}) overlaid on the PAndAS number density map.}
    \label{fig:age_metal_gc}
\end{figure}

Following the age classification of \citet{Usher2024MNRAS.528.6010U}, the GC sample is divided into three groups: old ($>$ 7~Gyr), intermediate (5–7~Gyr), and young ($<$ 5~Gyr). As shown in Fig. \ref{fig:age_metal_gc}, the intermediate and young populations are spatially confined to the disc. Notably, all GCs located at R$\rm_{proj} >$ 25~kpc are older than 7~Gyr. This suggests that the young, potentially metal-rich clusters are likely associated with a post-merger starburst within the disc, whereas the older GC population exhibits a more diverse range of properties consistent with a composite origin of in-situ and accreted components. 


To complement our investigation, we may classify the GC sample of \citet{Usher2024MNRAS.528.6010U} into three metallicity bins: metal-poor ([Fe/H] $<$ -2), intermediate (-2 $\leq$ [Fe/H] $\leq$  -1), and metal-rich ([Fe/H] $>$  -1). Figure~\ref{fig:age_metal_gc} presents the spatial distribution of these groups. However, interpreting these distributions is challenging; the presence of intermediate and even metal-rich clusters within the inner halo indicates a complex accretion history, rather than a simple radial gradient.

\begin{table*}[t] 
\caption{Properties of Association 2 GCs.}
    \centering
    \footnotesize
    
    \begin{tabular*}{\textwidth}{@{\extracolsep{\fill}} c c c c c c c}  
    \hline
    \noalign{\smallskip} 
    
    Name & $\alpha_{J2000.0}$ & $\delta_{J2000.0}$ & V$_{\text{helio}}$ & V$_{\text{M31corr}}$ & [Fe/H] [dex]& [Fe/H] [dex] \\ 
         & [h:m:s] & [$^{\circ}:':''$] & [$\rm km\,s^{-1}$] & [$\rm km\,s^{-1}$] & \citep{Sakari2022MNRAS.512.4819S}  & \citep{Usher2024MNRAS.528.6010U}  \\
    
    \noalign{\smallskip}
    \hline
    \noalign{\smallskip}

     G1 & 00:32:46.5 & +39:34:40 & -335 & -31 &  -& -0.75$_{0.07} ^{0.07}$\\
     \noalign{\smallskip}
     G2 & 00:33:33.7 & +39:31:18 & -352 & -49 & - & -1.92$_{0.06} ^{0.07}$\\
     \noalign{\smallskip}
     H2 & 00:28:03.2 & +40:02:55	 & -519 & -212 & -&- \\
     \noalign{\smallskip}
     H7 & 00:31:54.5 & +40:06:47 & -426 & -121 & -1.34~$\pm$~0.20  & -\\
     \noalign{\smallskip}
     H8 & 00:34:15.4 & +39:52:53 & -463 & -160 &- &-\\
     \noalign{\smallskip}
     PA18 & 00:28:23.2 & +39:55:04 & -551 & -245 &  - &- \\
    \noalign{\smallskip}
    PA19 & 00:30:12.2 & +39:50:59 & -544 & -239 & -  & -\\
    \noalign{\smallskip}
    PA21 & 00:31:27.5 & +39:32:21 & -600 & -296 &  - &- \\
    \noalign{\smallskip}
    PA22 & 00:32:08.3 & +40:37:31 & -437 & -132 &  -1.11~$\pm$~0.20 & -\\
    \noalign{\smallskip}
    PA23 & 00:33:14.1 & +39:35:15 & -476 & -172 &  - & -\\
    \noalign{\smallskip}
    \hline

    \end{tabular*}
    
    \tablefoot{
    Coordinates and velocities are obtained from \citetalias{Veljanoski2014MNRAS.442.2929V}.
    }
    \label{tab:ass2 parameters}
\end{table*}
\begin{table*}[t] 
\caption{Properties of Stream C/D GCs.}
    \centering
    \footnotesize
    
    \begin{tabular*}{\textwidth}{@{\extracolsep{\fill}} c c c c c c c}  
    \hline
    \noalign{\smallskip} 
    
    Name & $\alpha_{J2000.0}$ & $\delta_{J2000.0}$ & V$_{\text{helio}}$ & V$_{\text{M31corr}}$ & [Fe/H] [dex]& [Fe/H] [dex] \\ 
         & [h:m:s] & [$^{\circ}:':''$] & [$\rm km\,s^{-1}$] & [$\rm km\,s^{-1}$] & \citep{Sakari2022MNRAS.512.4819S}  &  \citep{Usher2024MNRAS.528.6010U} \\
    
    \noalign{\smallskip}
    \hline
    \noalign{\smallskip}

     B517 & 00:59:59.9 & +41:54:06 & -277	 & 16 & -1.36~$\pm$~0.21 & -1.44$_{0.06} ^{0.08}$ \\
     \noalign{\smallskip}
     H24 & 00:55:43.9	 & +42:46:15 & -121 & 175&-1.58~$\pm$~0.20 & -0.90$_{0.13} ^{0.14}$ \\
     \noalign{\smallskip}
     PA41 & 00:53:39.5 & +42:35:14 & -94 & 203 & -1.96~$\pm$~0.21&- \\
     \noalign{\smallskip}
     PA43 & 00:56:38.8 & +42:27:17 & -135 & 160 &  - &- \\
     \noalign{\smallskip}
     PA44 & 00:57:55.8 & +41:42:57 & -349 & -54 &-2.07~$\pm$~0.21 &-2.34$_{0.08} ^{0.08}$\\
     \noalign{\smallskip}
     PA45 & 00:58:37.9 & +41:57:11 & -135 & 159 & -  & -\\
    \noalign{\smallskip}
    PA46 & 00:58:56.3& +42:27:38 & -132 & 162 & -2.02~$\pm$~0.21  & -2.14$_{0.07} ^{0.07}$\\
    \noalign{\smallskip}
    PA47 & 00:59:04.7& +42:22:35 & -359 & -64 &  - & -\\
    \noalign{\smallskip}
    PA49 & 01:00:50.0& +42:18:13 & -240 &  53 &  - & -\\
    \noalign{\smallskip}
    
    \hline

    \end{tabular*}
    
    \tablefoot{
           Coordinates and velocities are obtained from \citetalias{Veljanoski2014MNRAS.442.2929V}.
    }
    \label{tab:stream CD parameters}
\end{table*}

\end{appendix}

\end{document}